\documentclass[%
 reprint,
 amsmath,amssymb,
aps,
]{revtex4-1}
\DeclareMathOperator{\Tr}{Tr}

\usepackage{float}
\usepackage{subfigure}
\usepackage[title]{appendix}
\usepackage{graphicx}
\usepackage{dcolumn}
\usepackage{bm}
\usepackage{hyperref}
\usepackage[mathlines]{lineno}

\begin{document}


\title{Dynamics of entanglement and state-space trajectories followed by a system of four-qubit in the presence of random telegraph noise: common environment (CE) versus independent environments (IEs)}

\author{L.T. Kenfack}
 \email{kenfacklionel300@gmail.com}

\author{M. Tchoffo}%
\email{mtchoffo2000@yahoo.fr}

\author{M.N. Jipdi}%

\author{G. C. Fuoukeng}%
\altaffiliation[Also at:\,\,]{Laboratoire de G\'enie des Mat\'eriaux, P\^ole Recherche-Innovation-Entrepreneuriat (PRIE), Institut Universitaire de la C\^ote, BP 3001 Douala, Cameroon.}
  
\author{L. C. Fai}%

\affiliation{%
Mesoscopic and Multilayer Structure Laboratory, Department of Physics, Faculty of Science, University of Dschang, PO Box: 67 Dschang, Cameroon.
}%

\date{\today}

\begin{abstract}
The paper investigates the dynamics of entanglement and explores some geometrical characteristics of the trajectories in state space, in four-qubit Greenberger-Horne-Zeilinger (GHZ) - and W-type states, coupled to common and independent classical random telegraph noise (RTN) sources. It is shown from numerical simulations that: (i) the dynamics of entanglement depends drastically not only on the input configuration of the qubits and the presence or absence of memory effects, but also on whether the qubits are coupled to the RTN in a CE or IEs; (ii) a considerable amount of entanglement can be indefinitely trapped when the qubits are embedded in a CE; (iii) the CE configuration preserve better the entanglement initially shared between the qubits than the IEs, however, for W-type states, there is a period of time and/or certain values of the purity for which, the opposite can be found. Thanks to results obtained in our earlier works on the three-qubit model, we are able to conclude that entanglement becomes more robustly protected from decay when the number of qubits of the system increases. Finally, we find that the trajectories in state space of the system quantified by the quantum Jensen Shannon divergence (QJSD) between the time-evolved states of the qubits and some reference states may be curvilinear or chaotic.
\end{abstract}

\maketitle


\section{INTRODUCTION}
Nowadays, it is well established that multi-partite entangled states are powerful and indispensable resources for emerging quantum technologies such as quantum communication \cite{1,2,3}, quantum computing \cite{4}, quantum metrology \cite{5} and quantum imaging \cite{6}. However, the ultimate threat to the reliable practical implementation of quantum technologies base on creation and manipulation of multi-partite entangled states is the phenomenon of decoherence which is due to the unavoidable interaction of the quantum system with its external environment (open quantum system). In fact, such an interaction, independently to the quantum or classical nature of the external environment, is very fatal to the survival of the amount of quantum entanglement between the different constituent parts of a multi-partite quantum system \cite{7,35}. More precisely, the phenomenon of decoherence results in the destruction of the typical quantum properties of the system, such as quantum entanglement and consequently, after a finite interaction time, the state of the system may in some cases undergoes a transition from an entangled to a separable state. This situation is referred as early-stage disentanglement or entanglement sudden death (ESD) \cite{15,16} and has been observed experimentally by Almeida \textit{et al.} \cite{16r}. It is worth noting that the action of an environment on a quantum system may also have some non-detrimental effects. In point of fact, it has been recognized in the recent years that certain environments due to their non-trivial spectral structures and memory effects \cite{10,11} can enable the revival of entanglement (the so-called non-Markovian environments). This phenomenon known as entanglement revival (ER), is of great relevance from the practical point of view because it prolongs the lifetime of quantum entanglement and consequently it usage time. Moreover, decoherence is very important for crucial issues such as the classical-quantum measurement problem or crossing. Therefore, the study of the effects of decoherence on entangled systems would be of great importance not only for understanding how the system loses entanglement to the environment, but also for searching effective strategies to control, protect and even recover the entanglement. 
\par
There are in general two main approaches to describe the dynamics of open quantum systems: in the first one, the quantum system and the environment may be looked as a single quantum system whose evolution is governed by a unique global unitary operator. Thus, the dynamics of the system is easily obtained by tracing out the environment's degrees of freedom. In the second approach, one may consider the open quantum system under the action of an external stochastic classical field; here the dynamics of the system is obtained by replacing the trace over the environment's degrees of freedom by the average over the different realizations of the stochastic field \cite{17r}. The former pertains to the so-called quantum-mechanical modelling of the system-environment (S-E) interaction and the latter to the classical-mechanical one. However, it has been shown that for certain S-E interactions a classical description can be found that is completely equivalent to the quantum description \cite{12,13}. 
\par
In the last few decades, both theoretical and experimental investigations of the time evolution of entanglement in bipartite qubit systems under the influence of different decoherence models have attracted the interest of many researchers \cite{18r,19r,20r,21r,22r,23r,24r,25r,26r,27r,28r,lo1,lo2,lo3,lo4,lo5,lo6,lo7,lo8,lo9,lo10,Ken1}. However, for multi-qubit systems, the study of the time evolution of entanglement is very constrained by the lack of computable entanglement measures. In fact, until the present, the quantification and characterization of the exact amount of entanglement between the different constituent parts of a multi-partite entangled quantum system remains a challenging task and has been calculated only for particular model of decoherence and particular quantum states \cite{X2}. Nevertheless, many different entanglement measures for multi-partite entangled quantum systems defined as the sum of the bipartite entanglement measures over all the possible bi-partitions of total quantum system have been proposed during the years \cite{X3,X4} (see also the review papers \cite{8,9} and the references therein) and some interesting results based on this strategy have been obtained \cite{29r,30r,31r,32r,33r,34r,35r,36r,37r,38r,39r,40r}. For instance, it has been found that for multi-qubit systems, the degree of entanglement robustness tends to increases or decreases with the increase of the number of qubits of the system \cite{17,35}. In particular, the dynamics of quantum correlations in terms of entanglement and discord in a physical model of three non-interacting qubits subject to a classical environmental noise has recently been investigated \cite{27,28}. In the present work, in order to have an insight on how the entanglement behaves in the model studied in the above mentioned works (that is, \cite{27,28}) when the number of qubits is increased, we extend this model from three to four qubits. Stated another way, the main purpose of this work is to explore whether the increasing of the number of qubits in such a model leads to the increase or decrease of the entanglement fragility.
\par
In quantum information theory and computation, the notion of distance between quantum states has received considerable attention since it is closely related to quantum entanglement. Indeed, the use of the distance between two quantum states as a geometrical measure of entanglement has been investigated by A. P.  Majtey \textit{et al.} in \cite{14}. However, the distance between two quantum states can also serve to characterize the trajectories of a quantum system in the state space. For instance, in \cite{36r}, A. P.  Majtey \textit{et al.} have studied some features of the state-space trajectories followed by four-qubit systems subject to different noisy channels by computing the distance between the initially pure state of the system and its final mixed state obtained at the end of the decoherence process. 
\par
In the present work, we intend to investigate the dynamics of entanglement and explore some geometrical characteristics of the trajectories in state space in a physical model consisting of a system of four-qubit GHZ- and W-type states, coupled to a classical random telegraph noise (RTN). In particular, two different configurations of qubit-environment (Q-E) interaction are analysed. In the first one, each qubit locally interacts with its environment (local or independent environments interaction) while in the second one, all the qubits are embedded in a unique common environment (non-local or common environment interaction). In this work, the entanglement evolution is quantified in terms of genuine multi-qubit negativity \cite{4,35r}, and by means of the concept of the lower bound to multi-qubit concurrence (LBC) introduced in Ref. \cite{19} by Li \textit{et al.} These quantifiers are then compared with the detection proficiency of suitable entanglement witnesses. On the other hand, in order to explore the trajectories of the decohered GHZ and W states, we compute the distance between these states and some reference states such as the initial and the maximally mixed states. As estimator of the distance between two quantum states, we adopt the quantum Jensen Shannon divergence (QJSD) \cite{21,R}.
\par
The paper is organized as follows. In section~\ref{s2}, we briefly review the estimators adopted in this work to quantify the entanglement and characterize the state-space trajectories. In section~\ref{s3}, we present the physical model studied in this paper. In section~\ref{s4}, we investigate the dynamics of entanglement and explore some features of the state-space trajectories followed by the system under the action of RTN. Finally, in Section~\ref{s5}, we close the paper with some concluding remarks. 

\section{Quantification and characterization of the multi-qubit entanglement and state-space trajectories  }\label{s2}
In this section, we briefly expose the estimators adopted in this work to quantify and characterize the entanglement and the state-space trajectories followed by the system.  Note that all the material presented in this section is already known in the literature.
\subsection{Multi-qubit entanglement}
In the following, we use three well-defined measures of entanglement to quantify both analytically and numerically the amount of entanglement between the different qubits of the system: the genuine multi-qubit negativity, the lower bound to concurrence (LBC) and the concept of entanglement witnesses for multi-qubit entangled systems respectively. 

\subsubsection{N-qubit negativity}
In agreement with previous investigations \cite{4,35,35r}, one of the most useful and practical measures proposed to quantify the global amount of genuine entanglement of an arbitrary N-qubit entangled system in a mixed state $\rho\equiv \rho_{1,2,\hdots,N} $ is given by the average of the bipartite entanglement measures over all the possible bi-partitions of the N-qubit system. Its mathematical definition can be expressed as:
\begin{equation}
\mathcal{N}^{(N)}(\rho)=\dfrac{2}{N}\sum\limits_{k=1}^{N/2}\left( \dfrac{1}{n_{bipart}^{(k)}}\sum\limits_{P=1}^{n_{bipart}^{(k)}}\mathcal{N}^{P[k|N-k]}(\rho)\right), 
\end{equation}
where $ k\arrowvert N-k $ represent the bi-partitions of the N-qubit system with $ k $ qubits in one block and the remaining $ N-k $ ones in another block. $ P\left[ k\arrowvert N-k\right]  $ is use to specify a precise combination of $ k $ and $ N-k $ qubits in constituting the bipartition $ k\arrowvert N-k $. Thus, $ n_{bipart}^{(k)} $ stands for the total possible non-equivalent concrete bi-partitions $ P\left[ k\arrowvert N-k\right] $. $ \mathcal{N}^{P[k|N-k]}(\rho) $ denotes the bipartite entanglement (in terms of negativity) associated to the concrete bipartition $ P\left[ k\arrowvert N-k\right] $. As mentioned above, the entanglement associated with any given bipartition of the N-qubit system is evaluated by means of negativity, defined for an arbitrary N-qubit system in a mixed state $ \rho $ as:
\begin{equation}
\mathcal{N}^{P[k|N-k]}(\rho)=\sum\limits_{\jmath}\lvert\lambda_{\jmath}(\rho^{T_{I}})\rvert-1.
\end{equation}
Where $ \lambda_{\jmath}(\rho^{T_{I}}) $ are the eigenvalues of the partial transpose $ \rho^{T_{I}} $ of the total density matrix  with respect to the subsystem $ I $ which is constituted by the $ k $ qubits of the given bipartition $ P\left[ k\arrowvert N-k\right] $.

\subsubsection{N-qubit concurrence }
Beside the genuine N-qubit negativity, another powerful entanglement measure for multi-qubit entangled systems is the multi-qubit concurrence. However, unlike the genuine N-qubit negativity its calculation for multi-qubit entangled systems in mixed states is a challenging task since it involves difficult optimization process, which is very hard to solve exactly. To this end, many different analytically computable lower bounds to concurrence (LBC) were recently proposed \cite{19,18,20}. In this work, we will use the LBC suggested by Li \textit{et al.} \cite{19}, which is defined for an arbitrary N-qubit system in a mixed state $\rho\equiv \rho_{1,2,\hdots,N} $ as:
\begin{equation}
\mathcal{C}^{(N)}(\rho)=\sqrt{\dfrac{1}{N}\sum\limits_{\jmath=1}^{N}\sum\limits_{\ell=1}^{L}\left(\mathcal{C}_{\ell}^{(\jmath)}(\rho)\right)^{2}}.
\end{equation}
It can be appreciated from this equation that $ \mathcal{C}^{(N)}(\rho) $ is given in terms of N bipartite concurrences $ \mathcal{C}^{\jmath}(\rho) $ that correspond to the possible bipartite cuts of the N-qubit system in which just one of the qubits is separated from the remaining N-1 qubits. The bipartite concurrence  $ \mathcal{C}^{\jmath}(\rho) $ for the separation of the $ \jmath^{th} $ qubit is defined by a sum of $ L=2^{N-2}(2^{N-1}-1) $ terms $ \mathcal{C}_{\ell}(\rho) $ which is expressed as:
\begin{equation}
\mathcal{C}_{\ell}^{\jmath}(\rho)=\max\left\lbrace 0,\lambda_{\ell}^{1}-\lambda_{\ell}^{2}-\lambda_{\ell}^{3}-\lambda_{\ell}^{4}\right\rbrace.
\end{equation}
Where $ \lambda_{\ell}^{m} $, with $ m=1,2,3,4 $ are the square roots of the four non-vanishing eigenvalues in decreasing order of the non-Hermitian matrix $ \widetilde{\rho}=\rho\left[ L_{\jmath}^{\ell}\otimes L_{0}\rho^{\ast}L_{\jmath}^{\ell}\otimes L_{0}\right] $, with $ L_{0} $ the generator of the group $ SO(2) $ and $ L_{\jmath}^{\ell} $ the $ L=2^{N-2}(2^{N-1}-1) $ generators of the group $ SO(2^{N-1}) $. 

\subsubsection{N-qubit entanglement witnesses}
As we have already mentioned in the introduction, another very useful tool for the analysis of multi-qubit entanglement both in theory and experiment is the so-called ``entanglement witnesses''.  An observable W is called an entanglement witness if it satisfies the following properties \cite{9}: $ \Tr(\mathcal{W}\rho_{s}) \geq 0$ for all separable state $ \rho_{s} $ and $ \Tr(\mathcal{W}\rho_{e})<0$ for at least one entangled state $ \rho_{e} $. Note that there exists for each entangled state an entanglement witness detecting it. For details about the major construction methods of entanglement witnesses, we refer to Ref.~\cite{9}. However, for N-qubit GHZ- and W-type states, the general expressions of the corresponding entanglement witnesses are:
\begin{equation}
\mathcal{W}_{GHZ}^{(N)}=\dfrac{1}{2}\mathbb{I}_{N}-\lvert GHZ_{N}\rangle\langle GHZ_{N}\rvert
\end{equation}
and 
\begin{equation}
\mathcal{W}_{W}^{(N)}=\dfrac{N-1}{N}\mathbb{I}_{N}-\lvert W_{N}\rangle\langle W_{N}\rvert.
\end{equation}
Where $ \mathbb{I}_{N} $ is an $ N\times N $ identity matrix. s expected according to the definition of an entanglement witness operator given above, negative expectation values of the entanglement witnesses $ \mathcal{W}_{GHZ}^{(N)} $ and  $ \mathcal{W}_{W}^{(N)} $ indicate the appearance of multi-qubit entanglement experimentally detectable in the system meanwhile zero or positive expectation values do not guarantee the absence of entanglement \cite{28}. 

\subsection{State-space trajectories}
As pointed out in the introduction, in order to characterize the trajectories followed by the time-evolved states of the system affected by the RTN, we compute the distance between these states and some reference states such as the initial and the maximally mixed ones. In quantum information theory, a variety of measures such as for instance the quantum Jensen Shannon divergence (QJSD) \cite{21,R} and the Hilbert-Schmidt distance \cite{22} are used to quantify the distance between two quantum states. Specifically, in this work we will employ the QJSD to quantify the distance between the time-evolved states of the system and the above mentioned reference states. As shown in Refs.~\cite{21,R} the QJSD between two density operators $ \rho_{1} $  and $ \rho_{2} $  is defined in terms of the relative entropy as:
\begin{equation}
\mathcal{D}_{JS}\left( \rho_{1}\rVert\rho_{2}\right)=\dfrac{1}{2}\left[ \mathcal{S}\left( \rho_{1}\rVert\dfrac{\rho_{1}+\rho_{2}}{2}\right) +\mathcal{S}\left( \rho_{2}\rVert\dfrac{\rho_{1}+\rho_{2}}{2}\right)\right].
\end{equation}
The above equation can also be rewritten in terms of the von Neumann entropy $ \mathcal{H}_{N}(\rho)=-\Tr\left( \rho\log _{2}\rho\right)  $ as:
\begin{equation}
\mathcal{D}_{JS}\left( \rho_{1}\rVert\rho_{2}\right)=\mathcal{H}_{N}\left( \dfrac{\rho_{1}+\rho_{2}}{2}\right)-\dfrac{1}{2}\mathcal{H}_{N}(\rho_{1})-\dfrac{1}{2}\mathcal{H}_{N}(\rho_{2}).
\end{equation}
It is worth nothing that the QJSD satisfies the following properties: it is always well defined, symmetric, positive definite and bounded, that is $ 0\leq \mathcal{D}_{JS}\left( \rho_{1}\rVert\rho_{2}\right)\leq 1 $. 

\section{The physical model and Hamiltonian}\label{s3}
\begin{figure}[H]
\centerline{
\begin{tabular}{|c|c|}
\hline 
\subfigure[]{\includegraphics[width=0.23\textwidth]{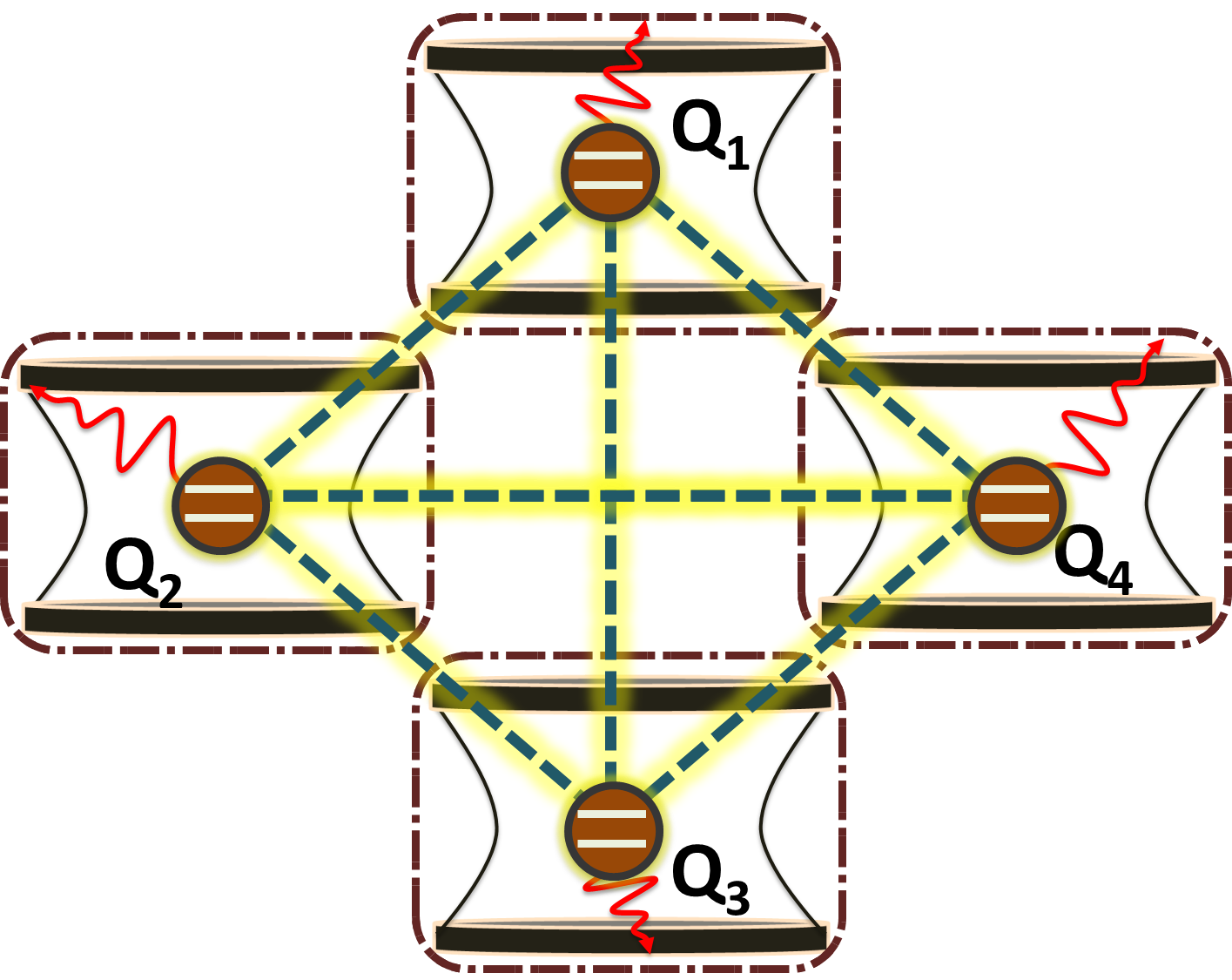}}
&
\subfigure[]{\includegraphics[width=0.23\textwidth]{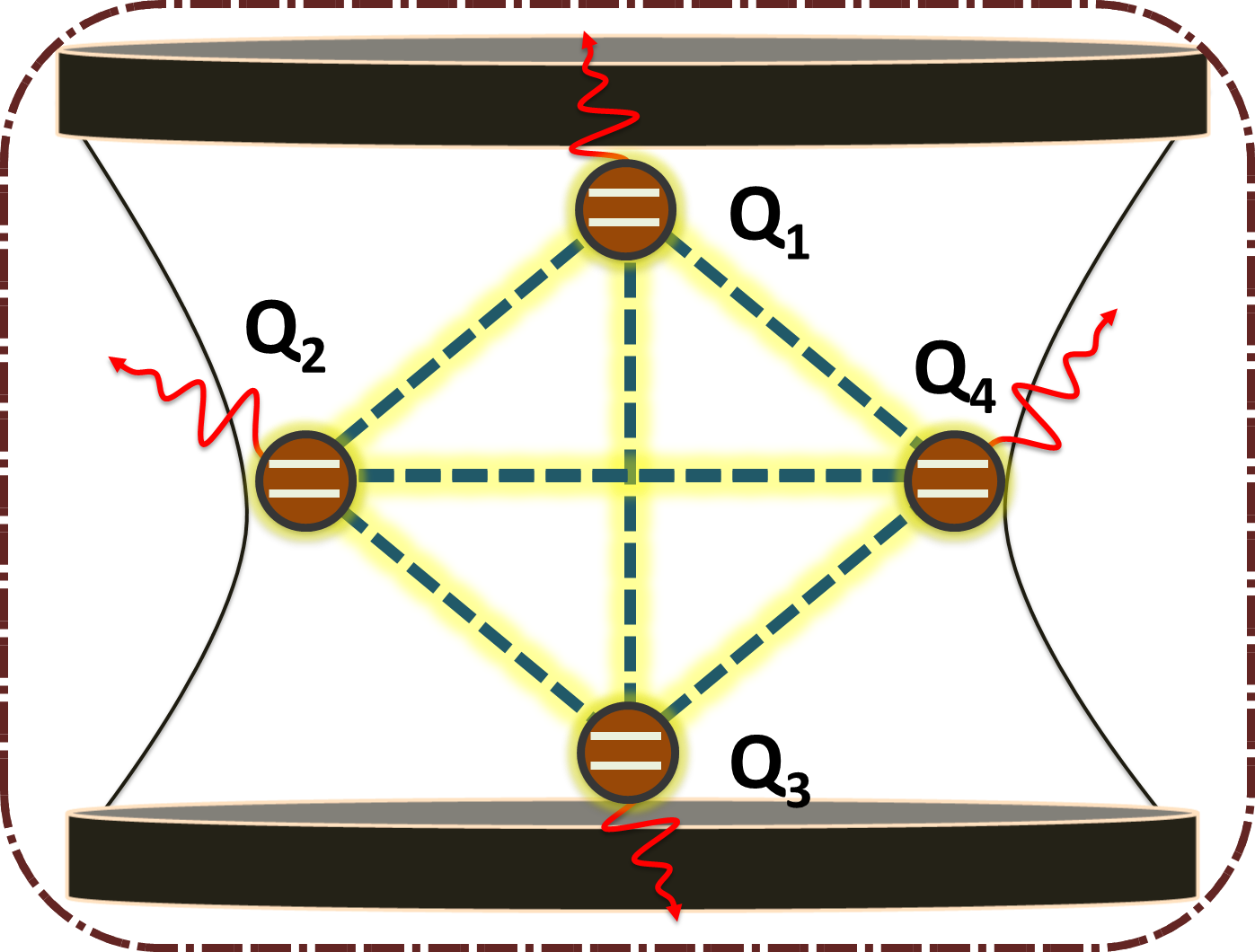} }
\\
\hline 
\end{tabular}} 
\caption{Schematic representations of the different Q-E coupling configurations studied in this paper: (a) local or independent environments coupling and (b) non-local or common environment coupling. The yellow dotted lines represent the entanglement initially shared between the qubits, while the red wavy lines show the interaction of each qubit with the classical environmental RTN source.}
\label{f1}
\end{figure}

The model studied in this work consists of four non-interacting qubits, initially entangled and subject to an environmental classical RTN in a common environment (CE) or in independent environments (IEs) as shown in Fig.~\ref{f1}. 
In the above Q-E coupling setups, the four qubits effective Hamiltonian can be written as follows:
\begin{equation}
\begin{split}
\mathcal{H}(t)&=\mathcal{H}_{1}(t)\otimes\mathbb{I}_{234}+\mathcal{H}_{2}(t)\otimes\mathbb{I}_{134}+\\&+\mathcal{H}_{3}(t)\otimes\mathbb{I}_{124}+\mathcal{H}_{4}(t)\otimes\mathbb{I}_{123},
\end{split}
\end{equation}
where $ \mathbb{I}_{JKL} $ represents the identity matrix acting on the Hilbert space of the qubits $ J $, $ K $ and $ L $ and $ \mathcal{H}_{Q}(t) $, $ Q=1,2,3,4 $ stands for the single qubit Hamiltonian which contains a stochastic term giving rise to the noise and can be explicitly expressed as:
\begin{equation}\label{3}
\mathcal{H}_{Q}(t)=\in_{0}\mathbb{I}_{Q}+\nu\vartheta_{Q}(t)\sigma_{Q}^{x}.
\end{equation}
$ \mathbb{I}_{Q} $ and $ \sigma_{Q}^{x} $ are respectively the identity operator and the spin-flip Pauli matrix acting on the Hilbert space of the $ Q^{th} $ qubit. $ \in_{0} $ is the qubit energy in the absence of noise (energy degeneracy is assumed), $ \nu $ is the strength of the Q-E coupling and $ \vartheta_{Q}(t) $ denotes a discrete stochastic term giving rise to the external RTN. To this end, the stochastic parameter $ \vartheta_{Q}(t) $ behaves as a bistable fluctuator \textit{i.e.,} it switches randomly between two values with a certain switching rate $ \gamma $ \cite{23}. The autocorrelation function of the stochastic parameter is an exponential decaying function given by:
\begin{equation}
\mathcal{K}_{RTN}(t-t')=\exp\left[ -2\gamma\lvert t-t'\rvert\right].
\end{equation}
However, when the qubits are coupled to the noise in a CE, we assume that $ \vartheta_{1}(t) =\vartheta_{2}(t)=\vartheta_{3}(t)=\vartheta_{4}(t)$  while when they are coupled to the noise in IE, we assume rather that $ \vartheta_{1}(t) \neq\vartheta_{2}(t)\neq\vartheta_{3}(t)\neq\vartheta_{4}(t)$. 
\par
\begin{figure}[H]
\centerline{\includegraphics[width=0.3\textwidth]{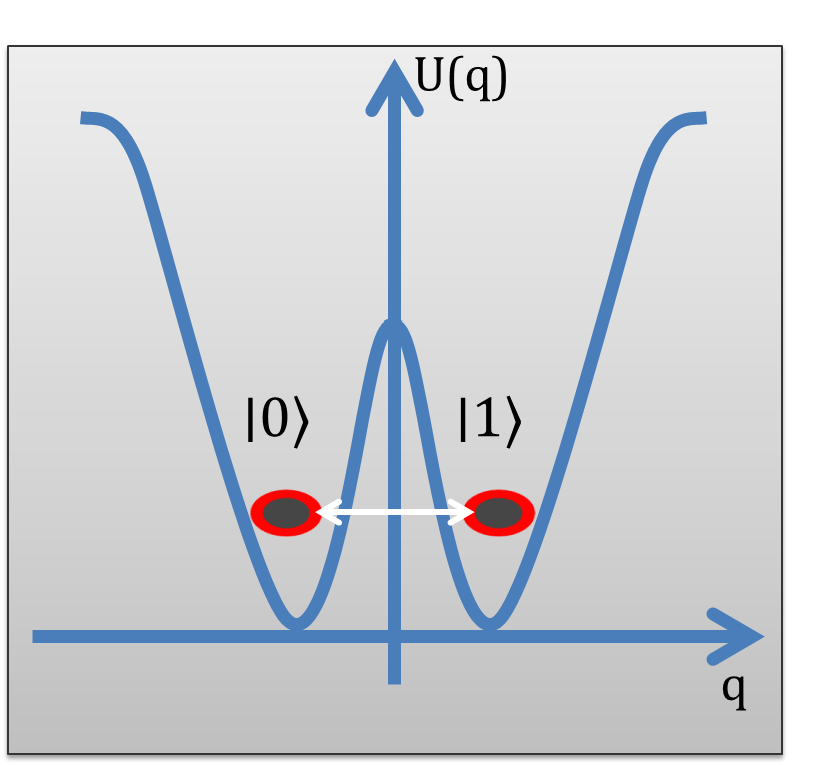}} 

\caption{Realization of a two-state system with degenerated energy levels, from the one dimensional spatial degrees of freedom of a single quantum particle trapped in a symmetric double-well potential. By tunnelling effect, the particle can be localized either in the right or in the left well. The state $ \lvert 0\rangle $ corresponds to situation when the particle is localized in the left well while the state $ |1\rangle $ corresponds to the case when the particle is localized in the right well. }
\label{f1a}
\end{figure}

The model Hamiltonian of Eq.~\eqref{3} describe an effective evolution of a quantum particle trapped in a symmetric double-well potential (see Fig.~\ref{f1a}), where the effects of noise is modelled by randomizing the tunnelling rate between the two wells. Note that, among the different kinds of qubit, our model is more linked to the Josephson vortex qubit \cite{l0}, which is fabricated by using a long Josephson junction ( which is a tri-stratum micro device in which a very slim insulator stratum is sandwiched between two relatively thick superconductor strata).

The similar Hamiltonian has been used to described four non-distinguishable quantum particles systems subjected to continuous-time quantum walks in a one-dimensional noisy lattice \cite{l1,l2} as well as weak coupling polarons in  spherical dot assisted with a transversal magnetic field \cite{l3}. Moreover, we note that its two- and three-qubit (particle) forms have recently been used to evaluate the dynamics of entanglement and quantum discord both analytically \cite{24,25,26,27,28,29,30,31} and numerically \cite{32,l1,l2}. Since there is not direct interaction between the four qubits, the evolution operator for the global system can be easily written as follows:
\begin{equation}
\mathcal{U}\left( \left\lbrace \vartheta\right\rbrace ,t\right)= \mathcal{U}_{1}\left(\vartheta_{1},t\right)\otimes\mathcal{U}_{2}\left(\vartheta_{2},t\right)\otimes\mathcal{U}_{3}\left(\vartheta_{3},t\right)\otimes\mathcal{U}_{4}\left(\vartheta_{4},t\right),
\end{equation} 
where $ \mathcal{U}_{Q}\left(\vartheta_{Q},t\right) $ stands for the single-qubit time-evolution operator and can be written in the units of $ \hslash=1 $ as:
\begin{equation}
\begin{split}
\mathcal{U}_{Q}\left(\vartheta_{Q},t\right)&=\exp{\left[-\imath\int\limits_{0}^{t}\mathcal{H}_{Q}(\tau)\,d\tau\right] }=\\&= 
e^{-\imath\in_{0}t}\left(\begin{matrix}
\cos(\eta_{Q}(t)) & -\imath\sin(\eta_{Q}(t)) \\ 
-\imath\sin(\eta_{Q}(t)) & \cos(\eta_{Q}(t))
\end{matrix} \right)
\end{split}
\end{equation}
where   $ \eta_{Q}(t)=\nu\int\limits_{0}^{t}\vartheta_{Q}(\tau)\,d\tau $ is the random noise phase picked up during the time interval $ \left[ 0,t\right]  $ and whose the explicit expression of the characteristic function is given by \cite{41r,42r}:
\begin{equation}
\begin{split}
&\Big\langle e^{\displaystyle\pm\imath\kappa\eta_{Q}(t)} \Big\rangle_{\eta_{Q}}=
\\&=
\left\lbrace
\begin{split}
&\displaystyle e^{\displaystyle-\gamma t}\left[ \cosh(\delta t)+\dfrac{\gamma}{\delta}\sinh(\delta t)\right] \rightarrow\,\gamma>\kappa\nu
\\\\&
\displaystyle e^{\displaystyle-\gamma t}\left[ \cos(\delta t)+\dfrac{\gamma}{\delta}\sin(\delta t)\right] \rightarrow\,\gamma<\kappa\nu
\end{split},
\right.
\end{split}
\end{equation}
where $ \delta=\sqrt{\lvert\gamma^{2}-(\kappa\nu)^{2}\rvert} $, with $ \kappa $ being an integer. Let us note that when $ \gamma>\kappa\nu $ the noise is Markovian (Markovian regime) meanwhile when $ \gamma<\kappa\nu $  the noise is rather Non-Markovian (Non-Markovian regime). If the qubits are initially prepared in a state $ \rho(0) $, the time evolution of the system under the influence of RTN is given by the averaged value of the evolved density matrix $ \rho\left(\left\lbrace \vartheta\right\rbrace,t\right)=\mathcal{U}\left( \left\lbrace \vartheta\right\rbrace,t\right)\rho(0)\mathcal{U}\left( \left\lbrace \vartheta\right\rbrace,t\right)^{\dagger}$ over all the possible realizations of the stochastic process $ \eta(t) $:
\begin{equation}\label{a}
\rho(t)=\Big\langle \mathcal{U}\left( \left\lbrace \vartheta\right\rbrace,t\right)\rho(0)\mathcal{U}\left( \left\lbrace \vartheta\right\rbrace,t\right)^{\dagger}\Big\rangle_{\left\lbrace \eta\right\rbrace },
\end{equation}
where $ \langle\hdots\rangle_{\left\lbrace \eta\right\rbrace } $ indicates the average over a given noise configuration  $ \left\lbrace \eta\right\rbrace=\left\lbrace \eta_{1},\eta_{2},\eta_{3},\eta_{4}\right\rbrace   $. In this work, we assume that the qubits are initially prepared in the four-qubit GHZ- and W-type states given respectively by:
\begin{equation}\label{5}
\rho_{GHZ_{4}}(0)=\dfrac{1-q}{16}\mathbb{I}_{16}+q\lvert GHZ_{4}\rangle\langle GHZ_{4}\rvert
\end{equation}
and
\begin{equation}\label{6}
\rho_{W_{4}}(0)=\dfrac{1-q}{16}\mathbb{I}_{16}+q\lvert W_{4}\rangle\langle W_{4}\rvert.
\end{equation}
The advantages of studying these states rely on the fact that they: (i) are maximally entangled, (ii) can be easily prepared and (iii) cannot be transformed into each other by any stochastic local operations assisted by classical communication (SLOCC) that is, both states do not belong in the same class of entanglement. In Eqs.~\ref{5} and \ref{6}, $ q $ stands for the purity of the initial state ranging from $ 0 $ to $ 1 $, $ \mathbb{I}_{16} $ is the $ 16\times 16 $ identity matrix, $ \lvert GHZ_{4}\rangle=\dfrac{1}{\sqrt{2}}\left( \lvert 0000\rangle+\lvert 1111\rangle\right)  $ and $ \lvert W_{4}\rangle=\dfrac{1}{2}\left( \lvert 0001\rangle+\lvert 0010\rangle+\lvert 0100\rangle+\lvert 1000\rangle\right)  $. Once the calculations are performed, the explicit forms of the final density matrices for both CE and IEs coupling are reported in the Appendices~\ref{A} and \ref{B} respectively. 

\section{Numerical simulation results and discussions}\label{s4}
In this section, we present the numerical simulation results of the time evolution of entanglement and explore some geometrical features of the state-space trajectories followed by a system of four non-interacting qubits interacting with a classical environmental RTN in common and independent environments.

\subsection{Entanglement}
\subsubsection{GHZ-type states: the case of Common environment coupling }
\begin{figure*}[]
\centerline{
\begin{tabular}{|c|c|c|}
\hline 
\subfigure[]{\includegraphics[width=0.25\textwidth]{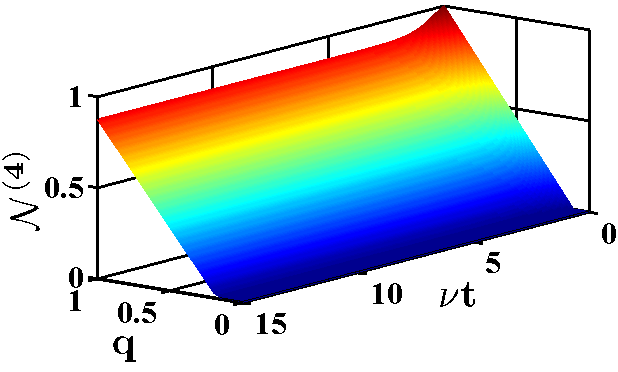}}&  
\subfigure[]{\includegraphics[width=0.25\textwidth]{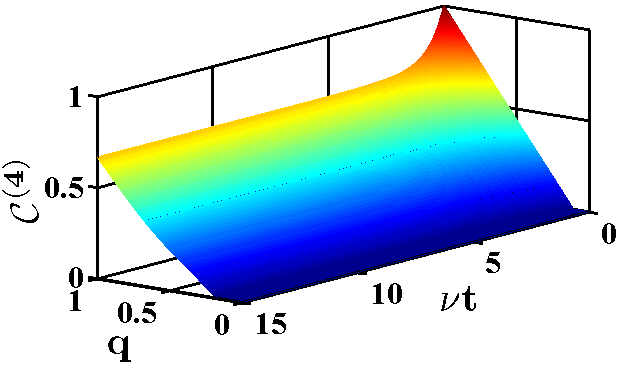}}& 
\subfigure[]{\includegraphics[width=0.25\textwidth]{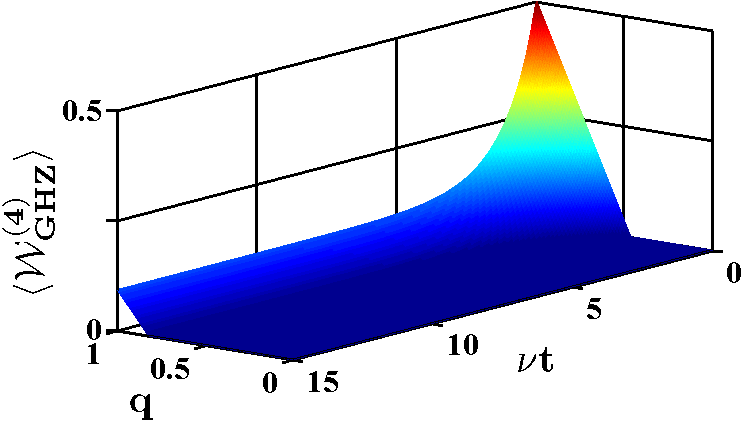}}\\ 
\hline 
\subfigure[]{\includegraphics[width=0.25\textwidth]{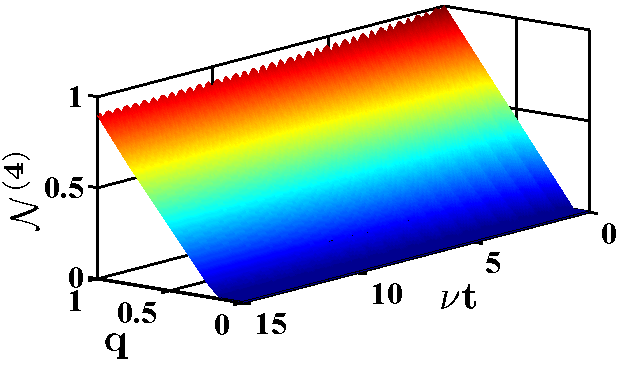}}&  
\subfigure[]{\includegraphics[width=0.25\textwidth]{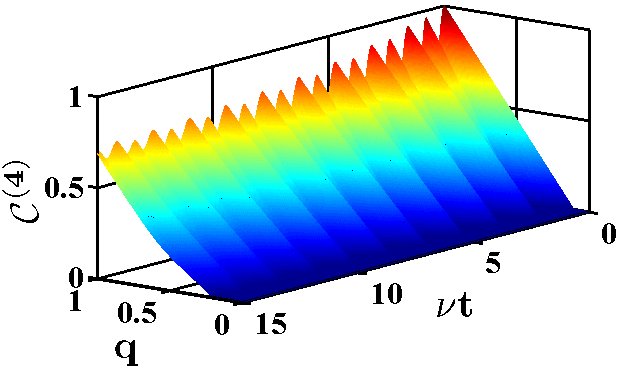}}& 
\subfigure[]{\includegraphics[width=0.25\textwidth]{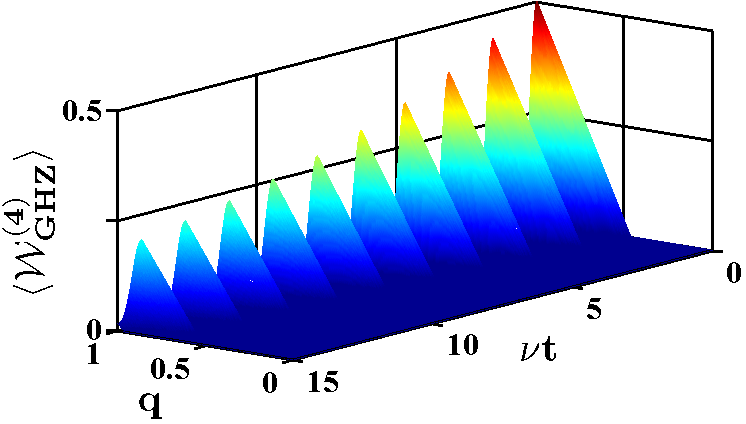}}\\ 
\hline 
\end{tabular}}
\caption{Upper panels: evolution of the four-qubit negativity $ \mathcal{N}^{(4)} $ (a), LBC $ \mathcal{C}^{(4)} $ (b) and the opposite of the expectation value of the $ \mathcal{W}_{GHZ}^{(4)} $ entanglement witness $ -\Big\langle\mathcal{W}_{GHZ}^{(4)}\Big\rangle $  (c) as a function of the dimensionless time $ \nu t $ and the purity $ q $ in the Markovian regime with $ \gamma/\nu=10 $, when the three qubits, initially prepared in the GHZ-type states of Eq.~\eqref{5} are coupled to the RTN in a common environment. Lower panels: same as in the upper panels in the non-Markovian regime with $ \gamma/\nu=0.1 $.}
\label{f2}
\end{figure*}

Here, the effects of RTN on the evolution of the entanglement of a four-qubit system are presented and discussed when the qubits are initially prepared in the GHZ-type states. The time evolution of the system for this input configuration in the case of CE coupling is reported in Eq.~\eqref{A1} of the Appendix~\ref{A}. Because of lack of compact expressions, the analytical results for the negativity and LBC are not presented here. However, the expectation value of the $ \mathcal{W}_{GHZ}^{(4)} $ entanglement witness can be expressed in terms of the function $ \beta_{\kappa}(t) $ (defined in Eq.~\ref{A3}) as:
\begin{equation}
\begin{split}
\Big\langle\mathcal{W}_{GHZ}^{(4)}\Big\rangle&=\Tr\left[\mathcal{W}_{GHZ}^{(4)}\rho_{GHZ_{4}}^{CE}(t)\right]=
\\&=\dfrac{7}{16}-\dfrac{3q}{8}\left[ \beta_{4}(t)+\dfrac{1}{12}\beta_{8}(t)+\dfrac{17}{12}\right].
\end{split}
\end{equation}
In Fig.~\ref{f2}, we report the evolution of the four-qubit negativity $ \mathcal{N}^{(4)} $, the concurrence $ \mathcal{C}^{(4)} $ and the opposite of the expectation value of the GHZ-type states witness's operator $ -\Big\langle\mathcal{W}_{GHZ}^{(4)}\Big\rangle $ as a function of the dimensionless time $ \nu t $ and the purity of the initial state in the Markovian $ \gamma/\nu =10$ and non-Markovian $ \gamma/\nu =0.1$ regime.
We observe that in the Markovian regime, the entanglement quantified in terms of $ \mathcal{N}^{(4)} $ and $ \mathcal{C}^{(4)} $ decays asymptotically until a given (finite) time $ \nu t $  and then freezes shortly to a stationary value while in the non-Markovian one, the entanglement decays with damped oscillations before reaching the stationary value. In interesting previous works \cite{27,28,26}, a similar behaviour was found in the three-qubit systems where entanglement of an initial GHZ state is also indefinitely preserves when the qubits are embedded in a common environment. However, the stationary value of the entanglement (\textit{i.e.,} the amount of the preserved entanglement) in the four-qubit model is considerably higher compared to the case of three-qubit model, demonstrating that the four-qubit GHZ-type states preserves more entanglement with respect to the three-qubit ones. This result turns out to be in good agreement with what is claimed in the literature that, in general, the entanglement exhibited by multi-qubit systems becomes more robust as the number of the qubits of the system increases \cite{17}. The survival of entanglement in the long-time limit has been ascribed to the indirect interaction between the qubits resulting from their coupling to a common source of noise \cite{28}. In point of fact, in Ref.~\cite{28}, the authors assert that when the qubits are embedded in a common environment, the latter can be interpreted as a sort of interaction conciliator between the qubits themselves. Such an interaction somehow not only counteracts the total suppression of entanglement at sufficiently long times, but also prevents its disappearance at finite times. More precisely, we find that the entanglement decays asymptotically with time until reaching the corresponding stationary value depending upon the purity $ q $ and stay constant along the dynamics (Markovian regime) or exhibit damped oscillations (non-Markovian regime). On the other hand, we find that the stationary value reached by entanglement quantified in terms of $ \mathcal{N}^{(4)} $ is higher, both in Markovian and non-Markovian regime than that of the entanglement quantified in terms of the LBC $ \mathcal{C}^{(4)} $. For an initial state with purity greater than $ 0.821 $, we find that the long-time entanglement protection can successfully be revealed by the $ \mathcal{W}_{GHZ}^{(4)} $ entanglement witness in the Markovian regime. However, this is no longer true in the non-Markovian regime since the expectation value of the $ \mathcal{W}_{GHZ}^{(4)} $ entanglement witness takes zero or positive values at finite times. Furthermore, we find that the amount of entanglement quantified in terms of negativity and LBC result to be higher than the one detected by the $ \mathcal{W}_{GHZ}^{(4)} $ entanglement witness in both regimes. 

\subsubsection{GHZ-type states: the case of independent environments coupling}
\begin{figure*}[]
\centerline{
\begin{tabular}{|c|c|c|}
\hline 
\subfigure[]{\includegraphics[width=0.25\textwidth]{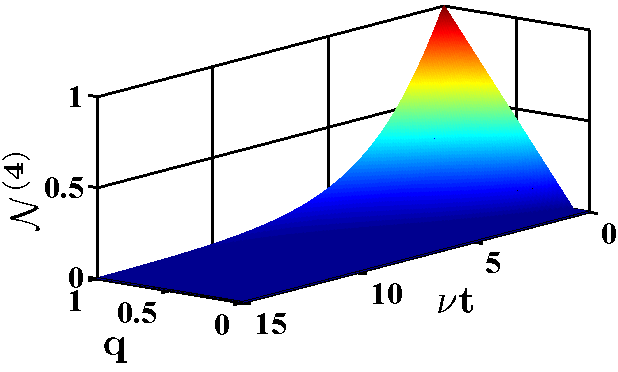}}&  
\subfigure[]{\includegraphics[width=0.25\textwidth]{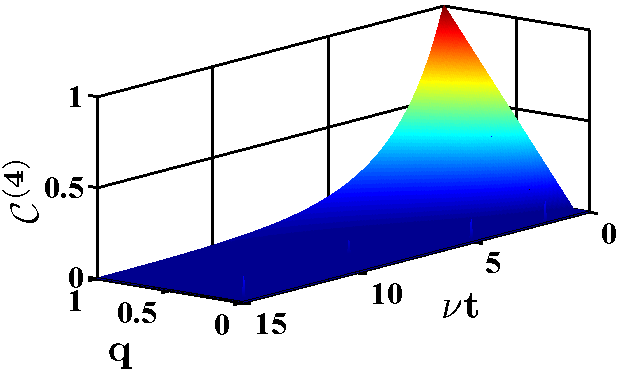}}& 
\subfigure[]{\includegraphics[width=0.25\textwidth]{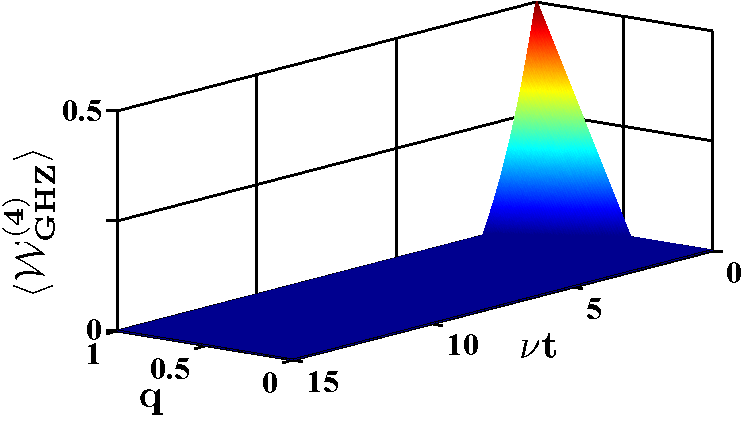}}\\ 
\hline 
\subfigure[]{\includegraphics[width=0.25\textwidth]{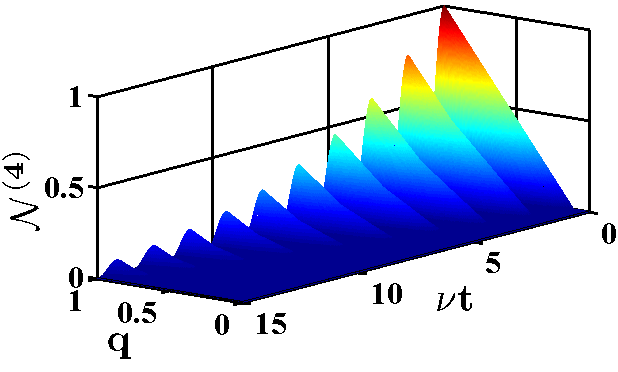}}&  
\subfigure[]{\includegraphics[width=0.25\textwidth]{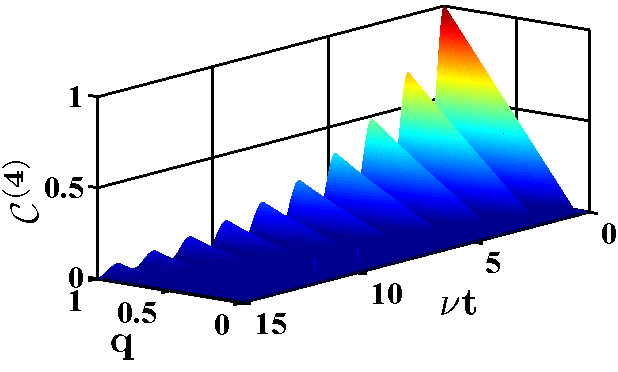}}& 
\subfigure[]{\includegraphics[width=0.25\textwidth]{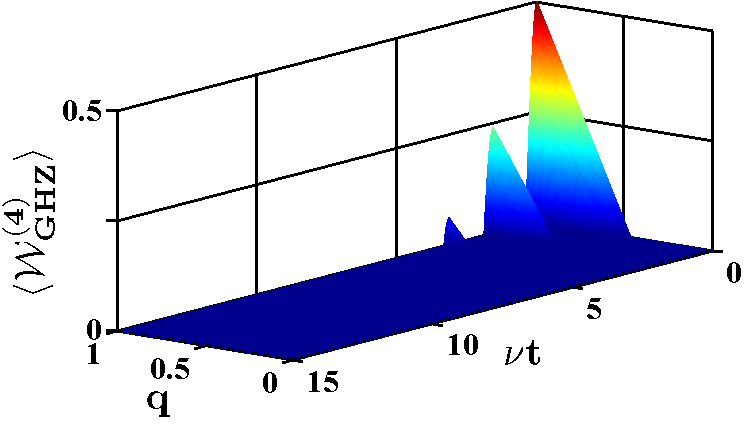}}\\ 
\hline 
\end{tabular}}
\caption{Upper panels: evolution of the four-qubit negativity $ \mathcal{N}^{(4)} $ (a), LBC $ \mathcal{C}^{(4)} $ (b) and the opposite of the expectation value of the $ \mathcal{W}_{GHZ}^{(4)} $ entanglement witness $ -\Big\langle\mathcal{W}_{GHZ}^{(4)}\Big\rangle $  (c) as a function of the dimensionless time $ \nu t $ and the purity $ q $ in the Markovian regime with $ \gamma/\nu=10 $, when the three qubits, initially prepared in the GHZ-type states of Eq.~\eqref{5} are coupled to the RTN in different environments. Lower panels: same as in the upper panels in the non-Markovian regime with $ \gamma/\nu=0.1 $.}
\label{f3}
\end{figure*}

Now, we investigated the effects of RTN on the evolution of entanglement when the four qubits, initially in the GHZ-type states of Eq.~\eqref{5} are coupled in different environments. For this configuration of the Q-E coupling, we find that the density matrix of the system at a given time t can be written as in Eq.~\eqref{A2} of the Appendix~\ref{A}. As we have already mentioned previously, only the numerical simulation results are presented because the negativity and the LBC cannot be put in a compact analytical form.  However, the expectation value of the $ \mathcal{W}_{GHZ}^{(4)} $ entanglement witness takes the following form:
\begin{equation}
\begin{split}
\Big\langle\mathcal{W}_{GHZ}^{(4)}\Big\rangle&=\Tr\left[\mathcal{W}_{GHZ}^{(4)}\rho_{GHZ_{4}}^{IE}(t)\right]=
\\&=\dfrac{7}{16}-\dfrac{q}{8}\left[ \beta_{2}^{4}(t)+6\beta_{2}^{2}(t)+\dfrac{1}{2}\right],
\end{split}
\end{equation}
where the function $ \beta_{k}(t) $  is defines as in Eq.~\eqref{A3}. In Fig.~\ref{f3}, we show the evolution of the four-qubit negativity $ \mathcal{N}^{(4)} $, the LBC $ \mathcal{C}^{(4)} $ and the opposite of the expectation value of the $ \mathcal{W}_{GHZ}^{(4)} $ entanglement witness $ -\Big\langle\mathcal{W}_{GHZ}^{(4)}\Big\rangle $ as a function of the dimensionless time $ \nu t $ and the initial purity of the state with $ \gamma/\nu=10 $ (Markovian regime) and $ \gamma/\nu=0.1 $ (non-Markovian regime). We observe that the entanglement decays asymptotically to zero and disappears completely after a given time in the Markovian regime while in the non-Markovian regime, it exhibits sudden death and revivals phenomena with damped amplitudes before disappears completely in the long-time limit. More precisely, in contrast to the case of CE coupling analyse previously, here, we find that entanglement is completely suppressed when the qubits are coupled to the RTN in different environments.
\par
Such a result is in good agreement with what has been found in the three-qubit model \cite{27,28} with the only difference that the entanglement dynamics of the four-qubit model can be robustly protected from decay than the one of the three-qubit model. The presence of revival phenomena is essentially due to the influence of the non-Markovian character or to the memory effects of the RTN which, in turn, because of the classical nature of the noise, is merely a consequence of the fact that knowledge of the state of the system at one time would provide information about which unitary trajectory (\textit{i.e.,} realization of the noise) the system is on, hence conditioning the future evolution. An overview on the phenomenon of entanglement revivals in classical environmental noise has been presented in \cite{33,34}. It follows from these works that the occurrence of entanglement revivals in a classical environment is the consequence of the fact that the environment (because of the presence of memory) keeps a classical record of what unitary operation has been apply to the system in such a way that if at certain time the environment loses the information about the system, the entanglement disappears and vice versa. On the other hand, we observe both in the Markovian and non-Markovian regime, that the initial amount of entanglement presents in the system as well as the amplitude of entanglement revivals decreases with the decrease of the purity of the initial state.  This, of course, can be explained by noticing that the decrease of the purity results in the increase of the degree of mixedness of the initial state. Moreover, at initial time, we find that for an initial state with purity less than $ 0.4667 $, the $ \mathcal{W}_{GHZ}^{(4)} $ entanglement witness is no longer able to detect the presence of entanglement. It is interesting to note that for three-qubit model, it has been shown that the presence of entanglement can be efficiently detected by the witness operator $ \mathcal{W}_{GHZ}^{(3)} $  only for states with purity higher than $ 0.4286 $. Furthermore, for equal values of the purity of the initial state, we remark that the curve generated by the LBC $ \mathcal{C}^{(4)} $  falls just slightly below that generated by the negativity $ \mathcal{N}^{(4)} $. In other words, we find that the survival time for $ \mathcal{N}^{(4)} $ is much longer than for the LBC $ \mathcal{C}^{(4)} $. It can be appreciated in Fig.~\ref{f3}(f) that the $ \mathcal{W}_{GHZ}^{(4)} $ entanglement witness is no longer able to detect the fourth and later entanglement revivals which is known to be present via the negativity and the LBC. This clearly demonstrate the weakness of the entanglement detected by means of witness over the one quantified by the negativity and LBC. Overall, we find that the action of CE or IEs coupling has different effects on the robustness of entanglement. In point of fact, we find that the CE coupling preserves better the entanglement between the qubits than the IEs coupling, regardless of the Markovian or non-Markovian character of the RTN. In other words, we find that the evolution of entanglement depends strongly on whether the qubits are coupled to RTN in a common environment or in independent environments.

\subsubsection{W-type states: the case of common environment coupling }

\begin{figure*}[]
\centerline{
\begin{tabular}{|c|c|c|}
\hline 
\subfigure[]{\includegraphics[width=0.25\textwidth]{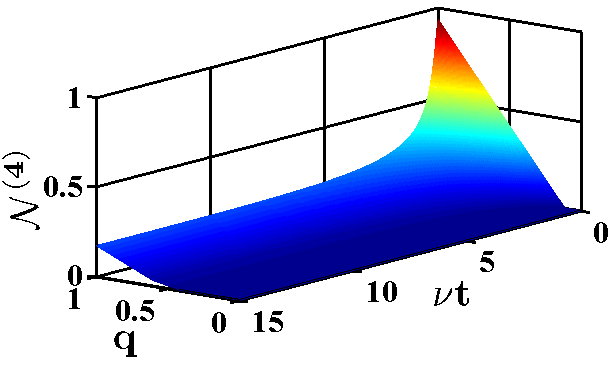}}&  
\subfigure[]{\includegraphics[width=0.25\textwidth]{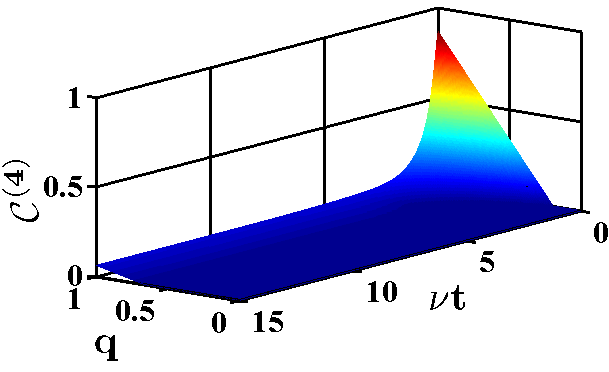}}& 
\subfigure[]{\includegraphics[width=0.25\textwidth]{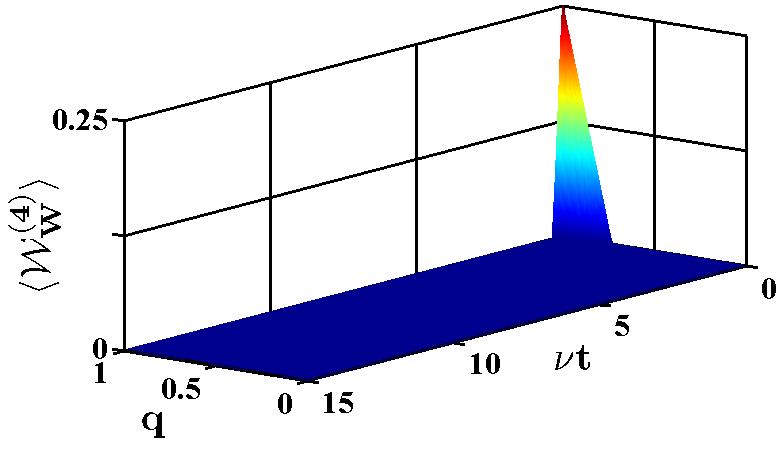}}\\ 
\hline 
\subfigure[]{\includegraphics[width=0.25\textwidth]{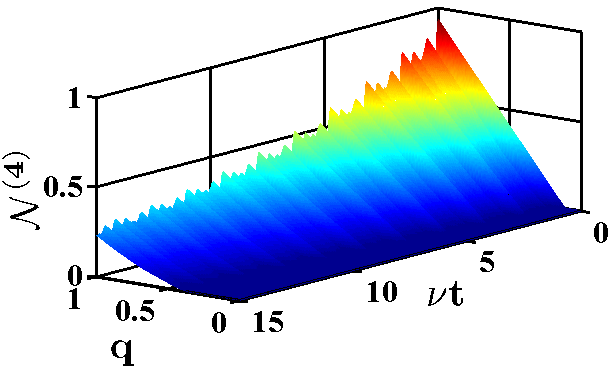}}&  
\subfigure[]{\includegraphics[width=0.25\textwidth]{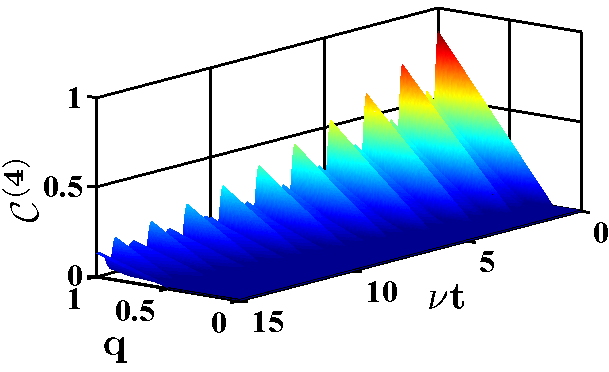}}& 
\subfigure[]{\includegraphics[width=0.25\textwidth]{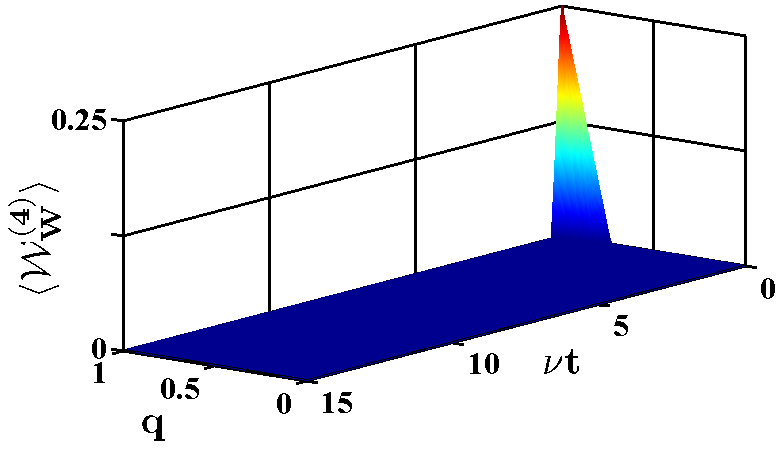}}\\ 
\hline 
\end{tabular}}
\caption{Upper panels: evolution of the four-qubit negativity $ \mathcal{N}^{(4)} $ (a), LBC $ \mathcal{C}^{(4)} $ (b) and the opposite of the expectation value of the $ \mathcal{W}_{W}^{(4)} $ entanglement witness $ -\Big\langle\mathcal{W}_{W}^{(4)}\Big\rangle $  (c) as a function of the dimensionless time $ \nu t $ and the purity $ q $ in the Markovian regime with $ \gamma/\nu=10 $, when the three qubits, initially prepared in the W-type states of Eq.~\eqref{6} are coupled to the RTN in a common environment. Lower panels: same as in the upper panels in the non-Markovian regime with $ \gamma/\nu=0.1 $.}
\label{f4}
\end{figure*}

Here, we examine the case in which the four qubit are initially prepared in the W-type states of Eq.~\eqref{6} and embedded in a common environment. After performing the calculations, we find that the final density matrix describing the evolution of the system can written as shown in Eq.~\eqref{B1} of the Appendix~\ref{B}. Once more, because of lack of compact analytical results, we deal only with the numerical ones. Nevertheless, the expectation value of the $ \mathcal{W}_{W}^{(4)} $ entanglement witness can expressed as:
\begin{equation}
\begin{split}
&\Big\langle\mathcal{W}_{W}^{(4)}\Big\rangle=\Tr\left[\mathcal{W}_{W}^{(4)}\rho_{W_{4}}^{CE}(t)\right]=
\\&=\dfrac{11}{16}-\dfrac{q}{4}\left[ \beta_{2}(t)+\dfrac{1}{2}\left( \beta_{4}(t)+\beta_{8}(t)\right)+\beta_{6}(t)+\dfrac{3}{4}\right]
\end{split}
\end{equation}
where the function $ \beta_{k}(t) $ is defines as in Eq.~\eqref{A3}. We report in Fig.~\ref{f4} the evolution of the four-qubit negativity $ \mathcal{N}^{(4)} $, the concurrence $ \mathcal{N}^{(4)} $  and the opposite of the expectation value of the W-type states entanglement witness's operator $ -\Big\langle\mathcal{W}_{W}^{(4)}\Big\rangle $ as a function of the dimensionless time $ \nu t $ and the purity of the initial state in the Markovian and non-Markovian regime.
\par 
For this input configuration, we find again that entanglement can be indefinitely preserved when the qubits are coupled in a common environment. Indeed, it can be appreciated in Fig.~\ref{f4} that for an initial state with purity ranging from $ 1 $ to $ 0.39 $, entanglement quantified in terms of negativity $ \mathcal{N}^{(4)} $ and LBC  $ \mathcal{C}^{(4)} $, decays asymptotically (Markovian regime) or with damped oscillations (non-Markovian regime) with time until reaching the corresponding stationary value depending upon the initial purity of the state. Note that the survival of entanglement in the long-time limits as shown by $ \mathcal{N}^{(4)} $ and $ \mathcal{C}^{(4)} $  represents the major discrepancy with what was found in the three-qubit form of the model studied in this paper \cite{27,28}. There, it has been shown that when the three qubits are initially set in a W-type state, entanglement is completely suppressed. This clearly indicated that the four-qubit W-type states preserve more entanglement than the three-qubit W-type ones. However, for an initial state with purity less than $ 0.39 $, the phenomena of ESD followed by ER (in the non-Markovian regime) appear. Moreover, we find that the $ \mathcal{W}_{W}^{(4)} $ entanglement witness fail to detect the presence of long-lived entanglement as well as the ER phenomena. Beyond this, we find that as the value of the purity $ q $ decreases in the initial state to $ 0 $, the non-Markovian behaviour becomes less evident. On the other hand, the entanglement detected by means of the $ \mathcal{W}_{W}^{(4)} $ entanglement witness demonstrates a similar behaviour in both regimes. As we have already pointed out in the case of GHZ-type states, for equal values of $ q $, the $ \mathcal{W}_{W}^{(4)} $ entanglement witness undergoes ESD before the negativity and the LBC, demonstrating that the entanglement measured by the negativity and the LBC is always higher with respect to the one detected by the witness. 

\subsubsection{W-type states: the case of independent environments coupling}
\begin{figure*}[]
\centerline{
\begin{tabular}{|c|c|c|}
\hline 
\subfigure[]{\includegraphics[width=0.25\textwidth]{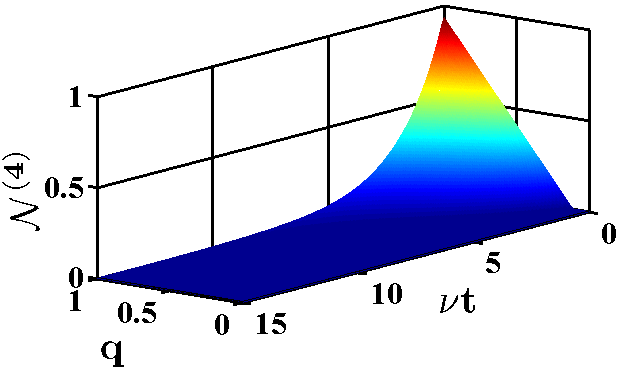}}&  
\subfigure[]{\includegraphics[width=0.25\textwidth]{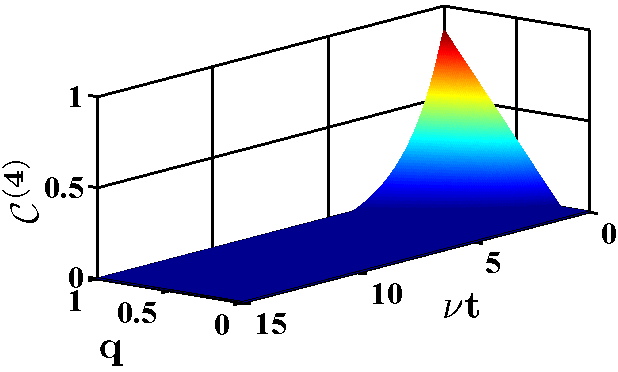}}& 
\subfigure[]{\includegraphics[width=0.25\textwidth]{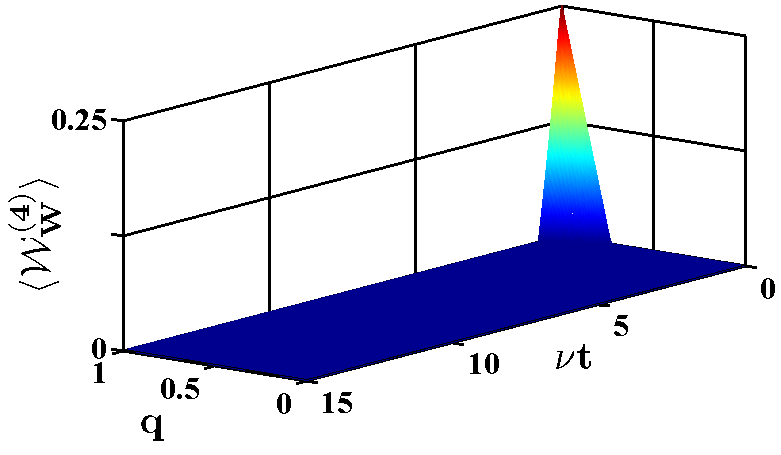}}\\ 
\hline 
\subfigure[]{\includegraphics[width=0.25\textwidth]{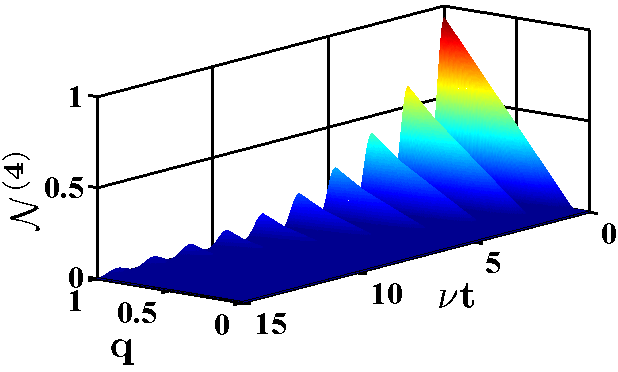}}&  
\subfigure[]{\includegraphics[width=0.25\textwidth]{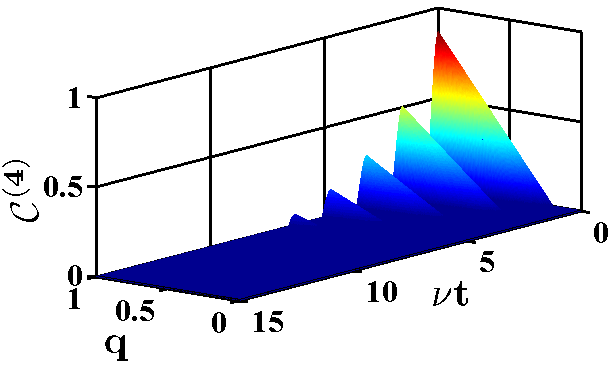}}& 
\subfigure[]{\includegraphics[width=0.25\textwidth]{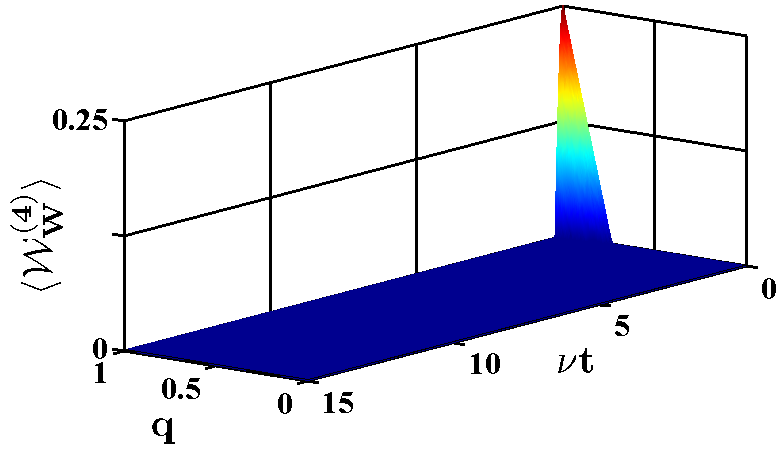}}\\ 
\hline 
\end{tabular}}
\caption{Upper panels: evolution of the four-qubit negativity $ \mathcal{N}^{(4)} $ (a), LBC $ \mathcal{C}^{(4)} $ (b) and the opposite of the expectation value of the $ \mathcal{W}_{W}^{(4)} $ entanglement witness $ -\Big\langle\mathcal{W}_{W}^{(4)}\Big\rangle $  (c) as a function of the dimensionless time $ \nu t $ and the purity $ q $ in the Markovian regime with $ \gamma/\nu=10 $, when the three qubits, initially prepared in the W-type states of Eq.~\eqref{6} are coupled to the RTN in different environments. Lower panels: same as in the upper panels in the non-Markovian regime with $ \gamma/\nu=0.1 $.}
\label{f5}
\end{figure*}
Finally, we analyse the time evolution of the entanglement when the four qubits, initially set in the W-type states of Eq.~\eqref{6}, interact with the RTN in independent environments. Once the calculations are performed, the state of the system for this configuration of the Q-E interaction, takes the form of Eq.~\eqref{B2} of the Appendix~\ref{B}. Following the definition of the $ \mathcal{W}_{W}^{(4)} $ entanglement witness, we can write the analytical form of its average value as:
\begin{equation}
\begin{split}
&\Big\langle\mathcal{W}_{W}^{(4)}\Big\rangle=\Tr\left[\mathcal{W}_{W}^{(4)}\rho_{W_{4}}^{IE}(t)\right]=
\\&=\dfrac{11}{16}-\dfrac{q}{4}\left[ \dfrac{5}{8}\beta_{2}^{4}(t)+\beta_{2}^{3}(t)+
\dfrac{3}{4}\beta_{2}^{2}(t)+\beta_{2}(t)+\dfrac{3}{8}\right]
\end{split}
\end{equation}
where the function $ \beta_{k}(t) $ is defines as in Eq.~\eqref{A3}. For this configuration, the dynamics of the four-qubit negativity  $ \mathcal{N}^{(4)} $, LBC $ \mathcal{C}^{(4)} $  and opposite of the expectation value of the $ \mathcal{W}_{W}^{(4)} $ entanglement witness  $ -\Big\langle\mathcal{W}_{W}^{(4)}\Big\rangle $ as a function of the dimensionless time and the purity for two different values of the ratio $ \gamma/\nu $ corresponding to Markovian (upper panels) and non-Markovian regime (lower panels) are illustrated in Fig.~\ref{f5}. As it can clearly be seen from this figure, the negativity and the LBC are monotonically deceasing function of time and vanish asymptotically in the Markovian regime; meanwhile in the non-Markovian regime they exhibit revival phenomena after their sudden death. Moreover, we observe that whatever the value of the purity is, the amount of entanglement quantified by the negativity is always higher than the one quantified by LBC which, in turn, is higher than the one detected by the $ \mathcal{W}_{W}^{(4)} $ entanglement witness. 
\par
At the initial time, we find that the $ \mathcal{W}_{W}^{(4)} $ entanglement witness detects the entanglement only when $ q>0.7333 $. This means that any initial state with purity less than $ 0.7333 $ cannot be any more detected by the witness. On the other hand, we also find that the phenomena of entanglement revival, which is known to be present in the non-Markovian regime via the negativity and the LBC, cannot be any more detected by the $ \mathcal{W}_{W}^{(4)} $ entanglement witness. Apart the fact that $ \mathcal{N}^{(4)} $  is higher than $ \mathcal{C}^{(4)} $ for equal values of the purity, we also find that $ \mathcal{N}^{(4)} $ exhibits more revival than $ \mathcal{C}^{(4)} $ and that at fixed value of time and for higher values of the purity, $ \mathcal{N}^{(4)} $ can be non-zero whereas $ \mathcal{C}^{(4)} $ is zero. Therefore, we are able to conclude that both measures are not compatible, that is, the negativity $ \mathcal{N}^{(4)} $ counts entanglement differently from the LBC $ \mathcal{C}^{(4)} $. 
\par
Overall, having compared the results obtained in this subsection with those obtained in the previous one, we find that the amount of entanglement initially present in the four-qubit GHZ-type states is more robustly shielded from decoherence than that of the four-qubit W-type states when the qubits are embedded in a common environment. This is in agreement with what was previously found in the three-qubit form of the model studied in the present paper \cite{25,27,28}. Besides, unlike GHZ-type states in which the CE coupling always preserves more entanglement than the IEs coupling, we find that for W-type states, the IEs and CE coupling may play opposite roles in the preservation of entanglement among the qubits, depending upon the purity $ q $ and the period of time considered. 

\subsection{Dynamics in the negativity-versus-mixedness space and influence of the switching rate in the degradation of entanglement}
\begin{figure}[b]
\centerline{
\begin{tabular}{|c|c|}
\hline  
\subfigure[]{\includegraphics[width=0.22\textwidth]{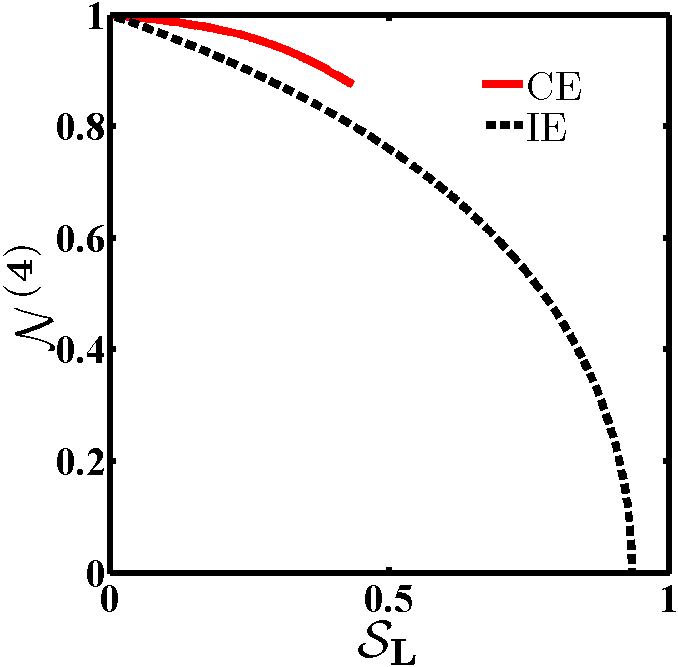}}&  
\subfigure[]{\includegraphics[width=0.22\textwidth]{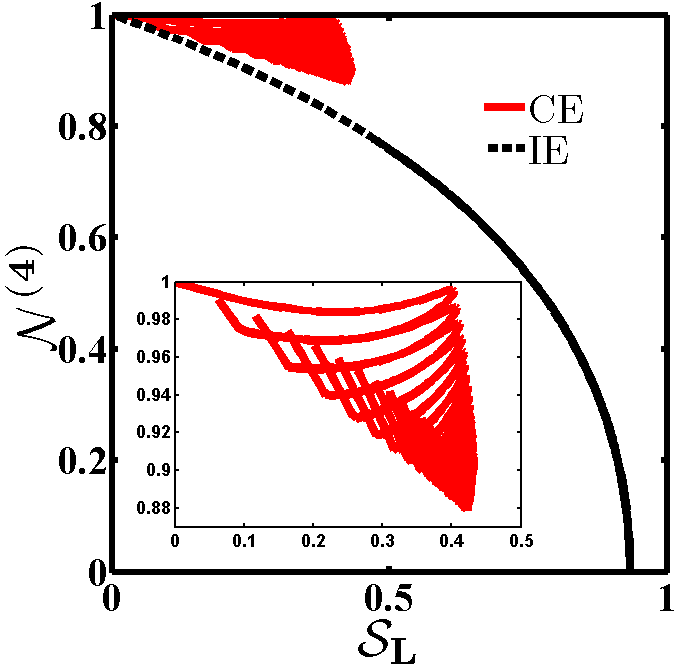}} \\ 
\hline  
\subfigure[]{\includegraphics[width=0.22\textwidth]{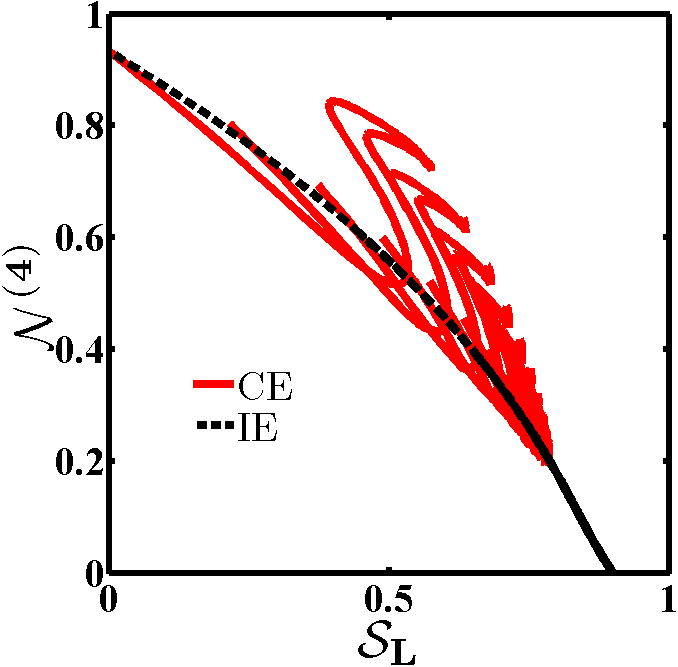}}&  
\subfigure[]{\includegraphics[width=0.22\textwidth]{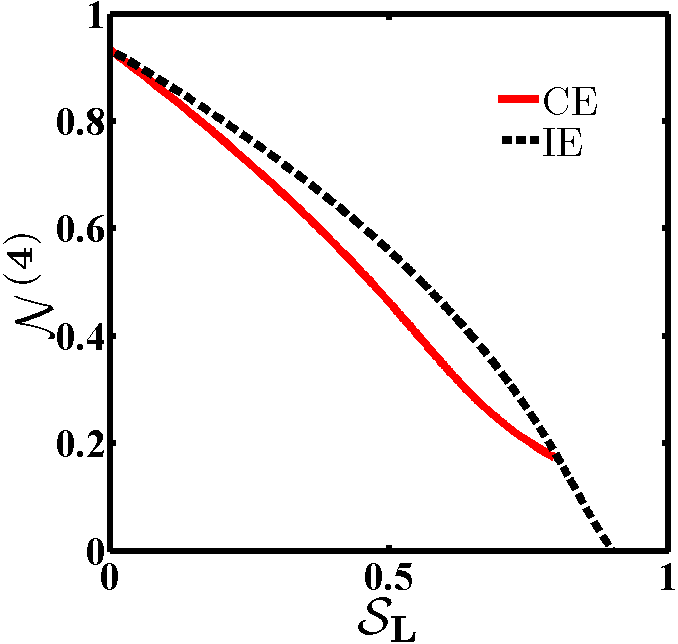}}\\ 
\hline 
\end{tabular}}
\caption{Upper panels: negativity versus linear entropy in the Markovian (a) and non-Markovian (b) regime for an initial pure GHZ state. Lower panels: same as in the upper panels for an initial pure W state.}
\label{f6}
\end{figure}

In order to have a better understanding on the relationship between the time evolution of entanglement of the initial pure GHZ and W state and their degree of mixedness, we investigate the interplay between the time evolution of entanglement and linear entropy. To this aim, we plot in Fig.~\ref{f6} the entanglement (quantified by the negativity) of the mixed states   $ \rho_{GHZ}(t) $ and $ \rho_{W}(t) $ obtained at each time step of the evolution of an initial pure GHZ and W state respectively, versus their degree of mixedness, given by the linear entropy
\begin{equation}
\mathcal{S}_{L}(\rho)=\dfrac{16}{15}\left[1-\Tr\left( \rho^{2}\right)  \right].
\end{equation} 
As expected, we observe that the negativity (and consequently the amount of entanglement) decreases with the increase of the linear entropy both in the Markovian and non-Markovian regimes. Concretely speaking, we observe that the amount of entanglement lost by the system due to its interaction with the RTN is accompanied by the loss of the purity of the state of the system, that is, the collapse of entanglement and purity occurs simultaneously. Therefore, one may conclude that the collapse of entanglement is strongly related to the transition of the state of the system from a pure quantum state to the corresponding mixed state. Evidently, for a given value of the linear entropy, we find that the pure GHZ state in the case of CE coupling (the broken curves) achieves the largest amount of negativity with respect to the IEs coupling, regardless of the Markovian or non-Markovian character the noise. However, the opposite is found for W state, where the residual amount of negativity can be smaller in the case of CE coupling than in the case of IEs coupling. It can also be appreciated in Fig.~\ref{f6} that some features of the entanglement evolution such as the long-time survival of entanglement the ESD and ER phenomena can be lighted up if we consider the trajectory followed by the system in the  $\mathcal{S}_{L}-\mathcal{N}^{(4)}  $ (mixedness-negativity) plane. Beyond this, we observe in the non-Markovian regime (in the case of CE coupling) that the trajectory of the system in the mixedness-negativity plane is chaotic both for GHZ and W state. This, of course, is consistent with the fact that all the quantities in the non-Markovian regime are oscillating functions of time and therefore, the negativity and the linear entropy tend to increase and decrease together. 
\par
\begin{figure}[b]
\centerline{
\begin{tabular}{|c|c|}
\hline  
\subfigure[]{\includegraphics[width=0.22\textwidth]{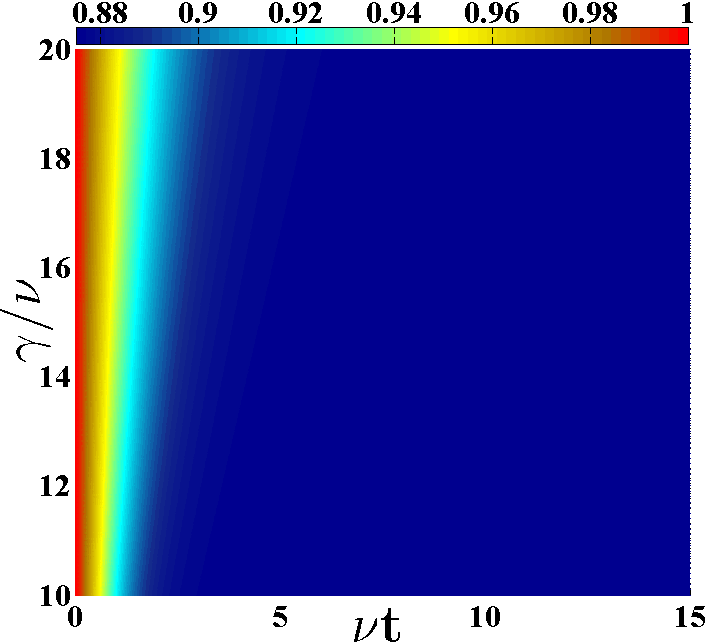}}&  
\subfigure[]{\includegraphics[width=0.22\textwidth]{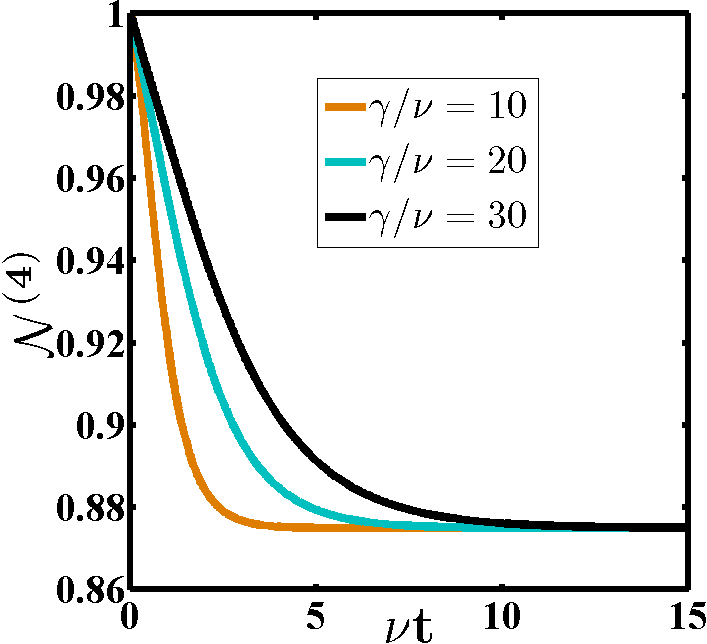}} \\ 
\hline  
\subfigure[]{\includegraphics[width=0.22\textwidth]{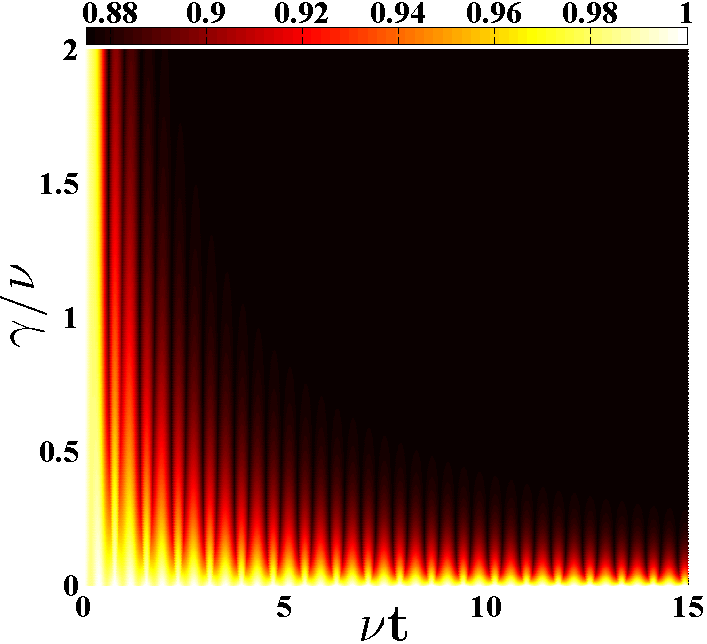}}&  
\subfigure[]{\includegraphics[width=0.22\textwidth]{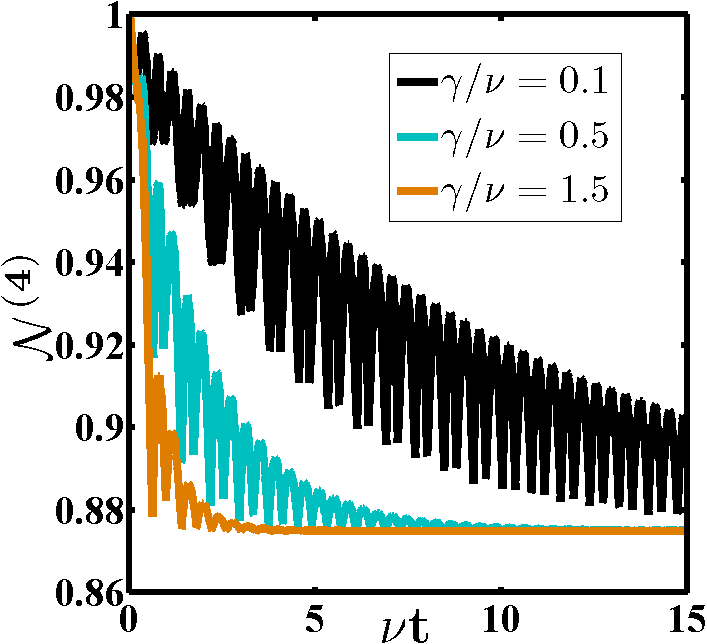}}\\ 
\hline 
\end{tabular}}
\caption{Upper panels: the contour plot of the negativity $ \mathcal{N}^{(4)} $ as function of the scaled switching rate  $ \gamma/\nu $ and time $ \nu t $  (a) and the corresponding plot for different values of the scaled switching rate (b) in the markovian regime in the case when the qubits are set in the GHZ state and embedded in a common environment.  Lower panels: same as in the upper panels but for non-Markovian regime.}
\label{f7}
\end{figure}
\begin{figure}[]
\centerline{
\begin{tabular}{|c|c|}
\hline  
\subfigure[]{\includegraphics[width=0.22\textwidth]{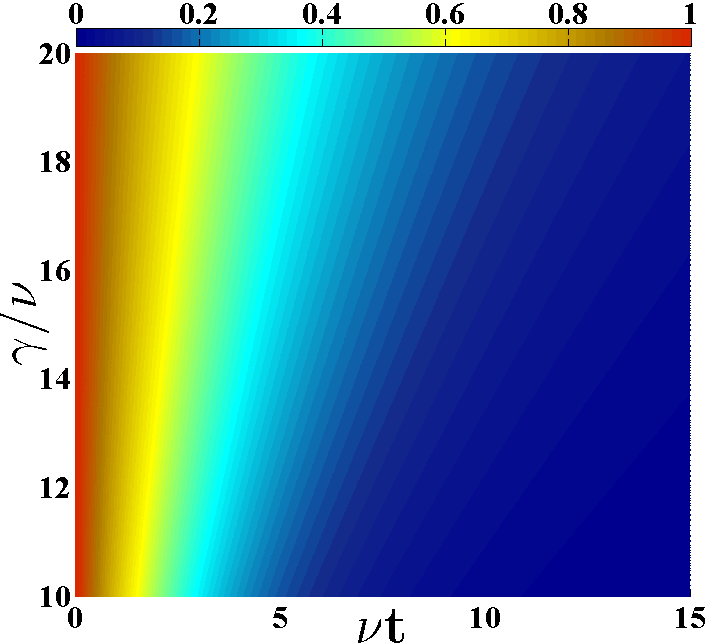}}&  
\subfigure[]{\includegraphics[width=0.22\textwidth]{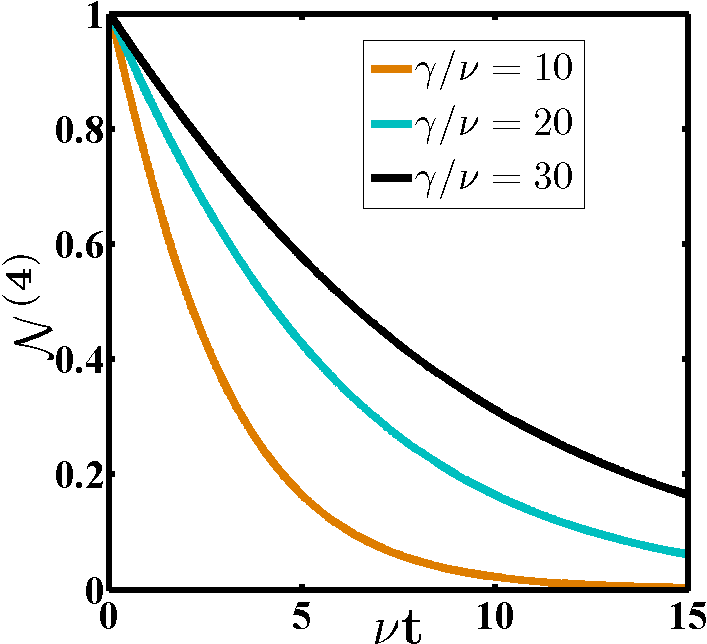}} \\ 
\hline  
\subfigure[]{\includegraphics[width=0.22\textwidth]{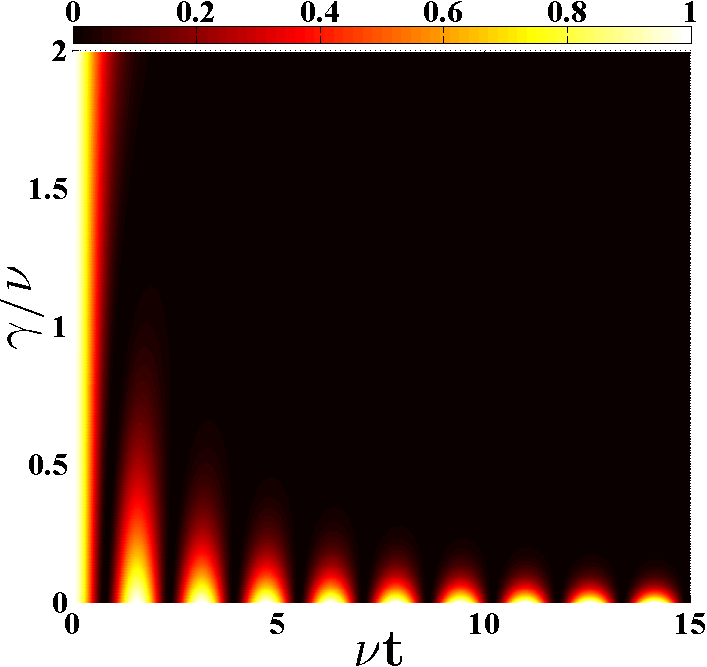}}&  
\subfigure[]{\includegraphics[width=0.22\textwidth]{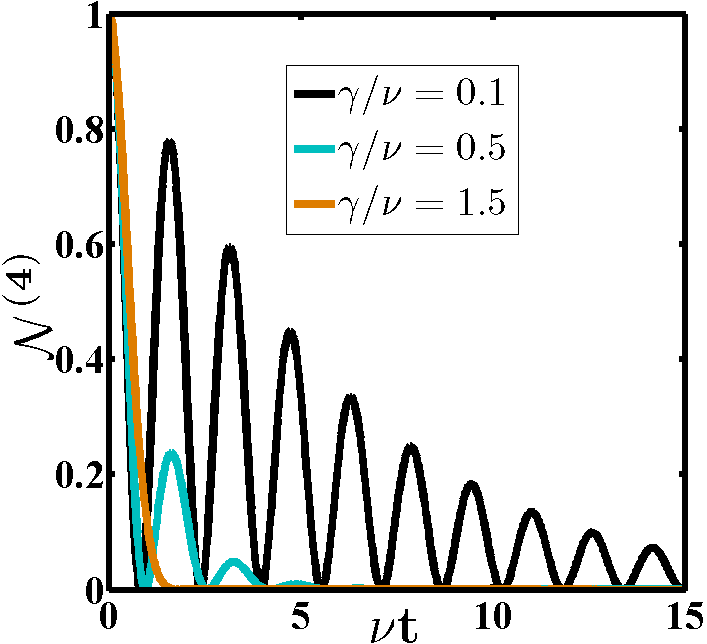}}\\ 
\hline 
\end{tabular}}
\caption{Same as in Fig.~\ref{f7} but in the case of independent environments.}
\label{f8}
\end{figure}

In the following we investigate the role played by the switching rate on the dynamics of entanglement. Figures~\ref{f7} and \ref{f8} show the evolutions of the four-qubit negativity as a function of the dimensionless switching rate $ \gamma/\nu $  and time $ \nu t $ , for qubits initially prepared in a pure GHZ state in the case of CE and IEs coupling respectively. 

\begin{figure}[]
\centerline{
\begin{tabular}{|c|c|}
\hline  
\subfigure[]{\includegraphics[width=0.22\textwidth]{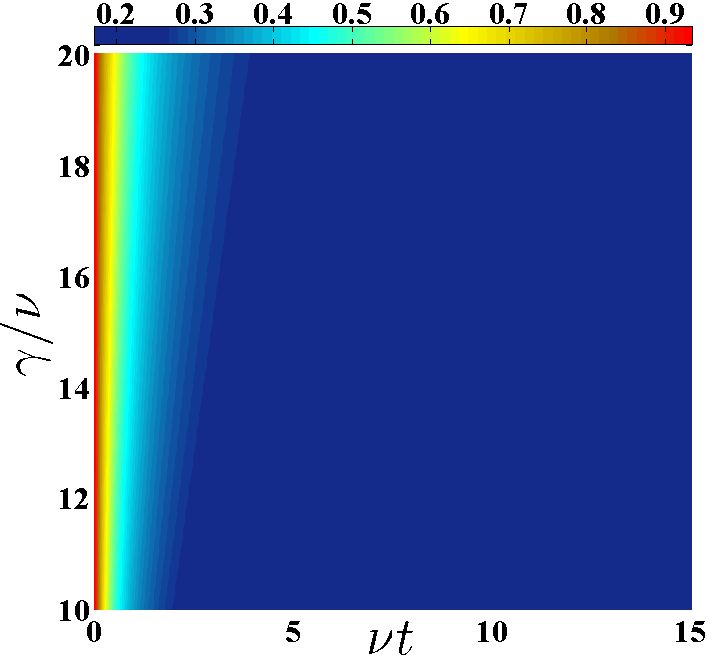}}&  
\subfigure[]{\includegraphics[width=0.22\textwidth]{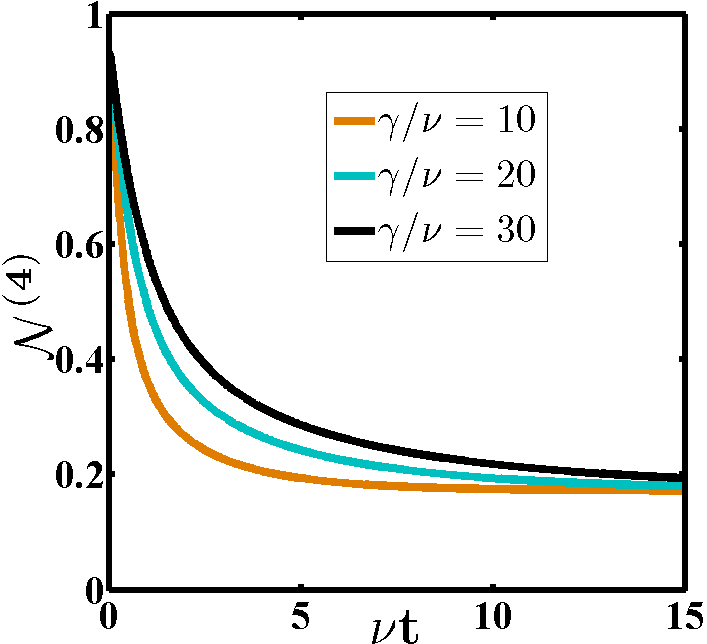}} \\ 
\hline  
\subfigure[]{\includegraphics[width=0.22\textwidth]{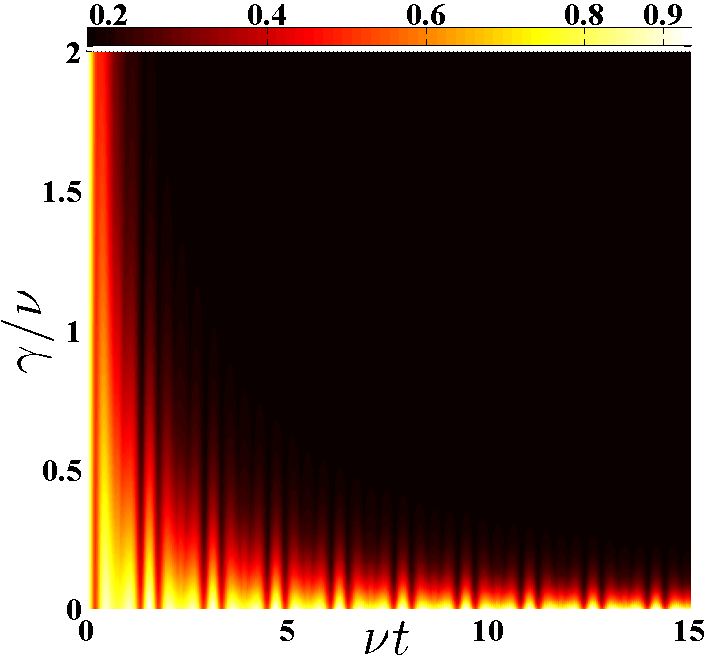}}&  
\subfigure[]{\includegraphics[width=0.22\textwidth]{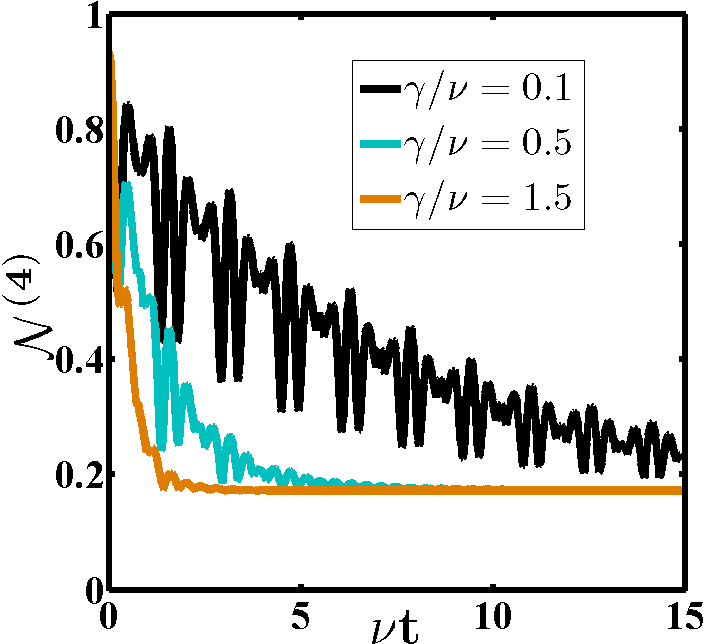}}\\ 
\hline 
\end{tabular}}
\caption{Upper panels: panel (a) shows the contour plot of the negativity as a function of the scaled switching rate $ \gamma/\nu $ and time meanwhile Panel (b) shows the dynamics of negativity for different values of the switching rate, both corresponding to the case when the qubits are coupled to the Markovian RTN in a common environment. Lower panels: same as in the upper one but when the qubits are coupled to the non-Markovian RTN.}
\label{f9}
\end{figure}

\begin{figure}[]
\centerline{
\begin{tabular}{|c|c|}
\hline  
\subfigure[]{\includegraphics[width=0.22\textwidth]{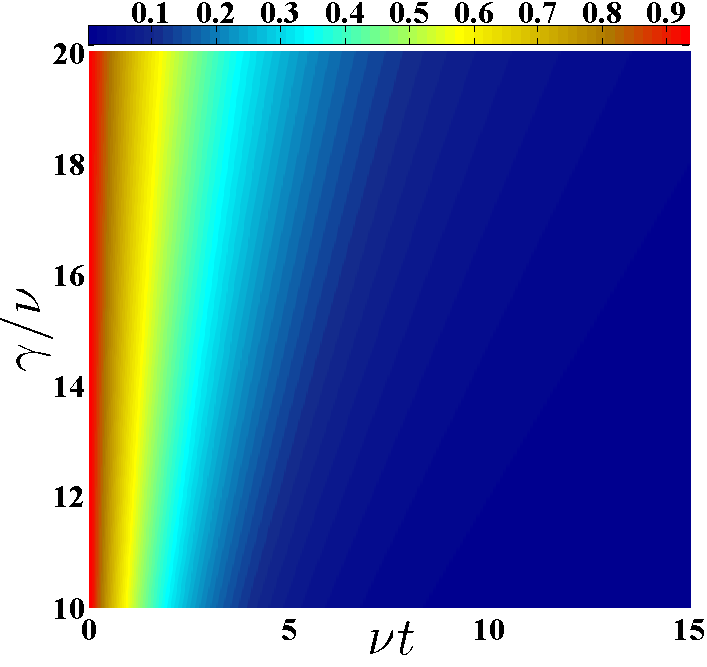}}&  
\subfigure[]{\includegraphics[width=0.22\textwidth]{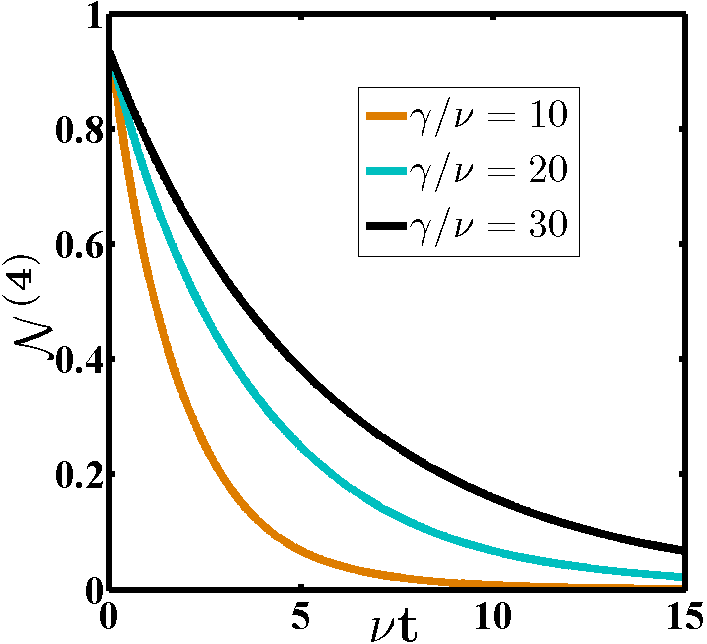}} \\ 
\hline  
\subfigure[]{\includegraphics[width=0.22\textwidth]{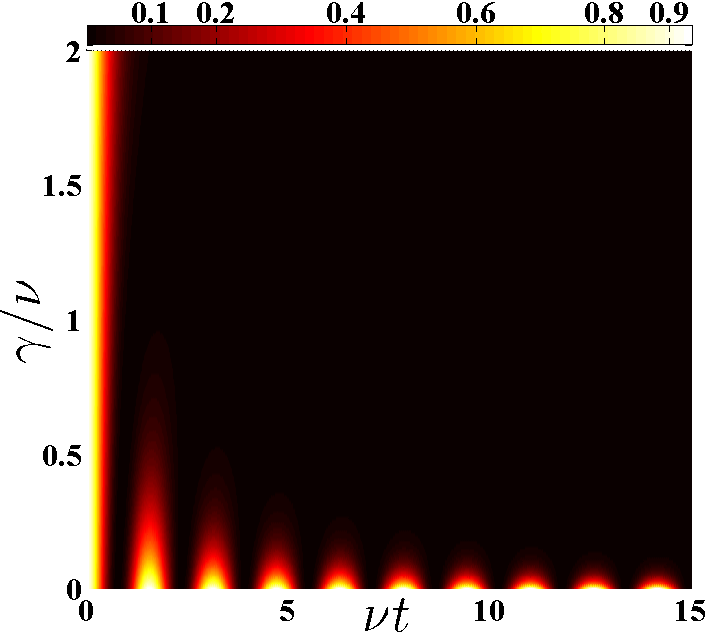}}&  
\subfigure[]{\includegraphics[width=0.22\textwidth]{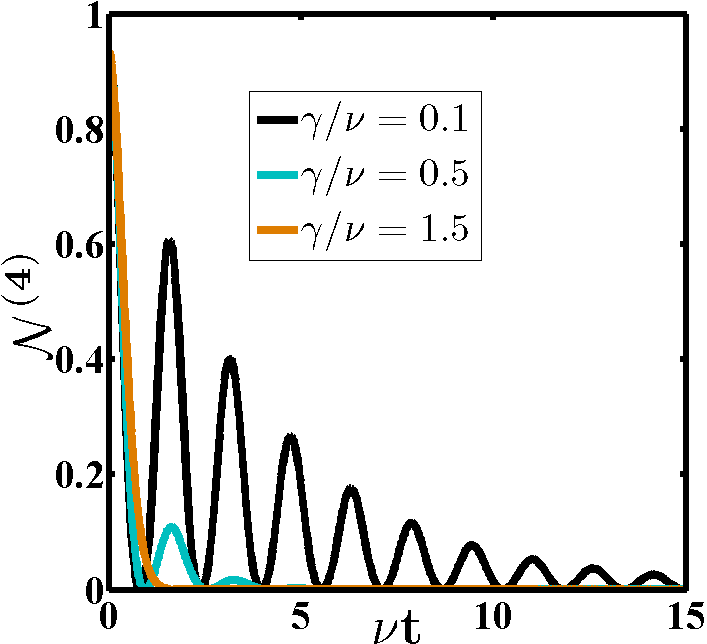}}\\ 
\hline 
\end{tabular}}
\caption{Same as in Fig.~\ref{f9} but when the qubits are coupled to the noise in independent environments. }
\label{f10}
\end{figure}

As expected according to the results obtained in the previous sections, when the qubits are embedded in a common environment, the entanglement does not any more disappear at a finite time and reaches a stable value shortly after the decay. In both figures one observes that if the qubits are weakly coupled to the external noise, that is, in the Markovian regime, the decay rate of entanglement can be greatly inhibited by increasing the switching rate. In other words, we observe that the larger is the value of the switching rate, the larger is the entanglement disappearance time. However, the opposite scenario can be observe in case of strong qubit-environment coupling (non-Markov regime). In point of fact, contrarily to the Markovian regime, we observe that the increasing of the switching rate results in a faster decay of entanglement in the non-Markovian one. These behaviours can be explained by noticing that smaller switching rates correspond to strong couplings between the qubits and the external environment while larger ones correspond to weak qubit-environment couplings. As a consequence, contrarily to the former, in the latter, more time is required for the system to endure the detrimental effects of the environmental noise. On the other hand, we mentioned once again that the independent environment acts as an entanglement decay among the qubits catalyst in the sense that its effect is more fatal to the survival of entanglement than that of a common environment acting on the qubits. Concretely speaking, one observes that when the qubits are embedded in a common environment, a relatively large amount of entanglement can be trapped in the system even at sufficiently long but finite time meanwhile entanglement is completely suppressed when we consider independent environments acting on each qubit. 
We now discus the effects of the switching rate on the dynamics of entanglement in the situation when the qubits are prepared in the pure W state. In Figs.~\ref{f9} and \ref{f10} we display the density plots of the entanglement as function of the rescaled switching rate $ \gamma/\nu $ and time  $ \nu t $ as well as the corresponding 2D plots for different values of the rate for both independent and common environment couplings in the Markovian and non-Markovian regime.

We can readily see from Figs.~\ref{f9} and \ref{f10} that the effects of the switching rate on the evolution of entanglement when the qubits are initially set in the W state are very similar to those obtained in  Figs.~\ref{f7} and \ref{f8} when the qubits are initially in the GHZ state. So, some conclusions are also similar and we are not going to repeat ourselves here. However, according to the results obtained above for GHZ state, we are able to assert that the amount of preserved entanglement depends on the choice of the input state of the qubits. Of course, in agreement with previous results, the preserved amount or steady value of entanglement is considerable higher for GHZ state than for W one, demonstrating that the former is more robust under the classical RTN than the latter when the qubits are embedded in a common environment. This, of course, is still true in the case of independent environments acting on each qubit even though in both input states, the entanglement disappears completely in the limit of long-time. 

\subsection{State-space trajectories}
\begin{figure}[]
\centerline{
\begin{tabular}{|c|c|}
\hline  
\subfigure[]{\includegraphics[width=0.22\textwidth]{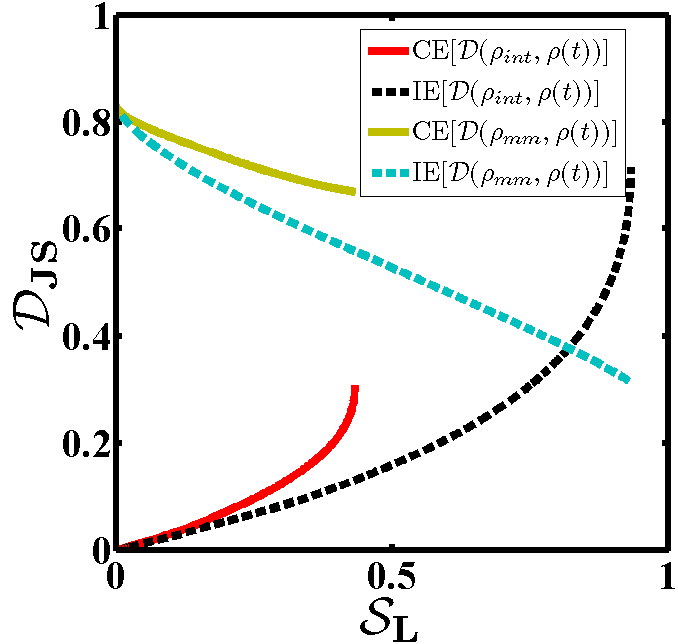}}&  
\subfigure[]{\includegraphics[width=0.22\textwidth]{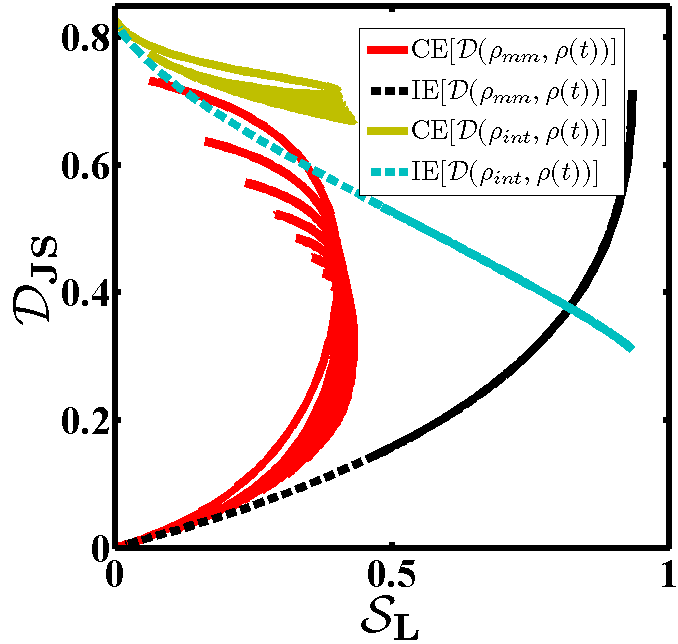}} \\ 
\hline  
\subfigure[]{\includegraphics[width=0.22\textwidth]{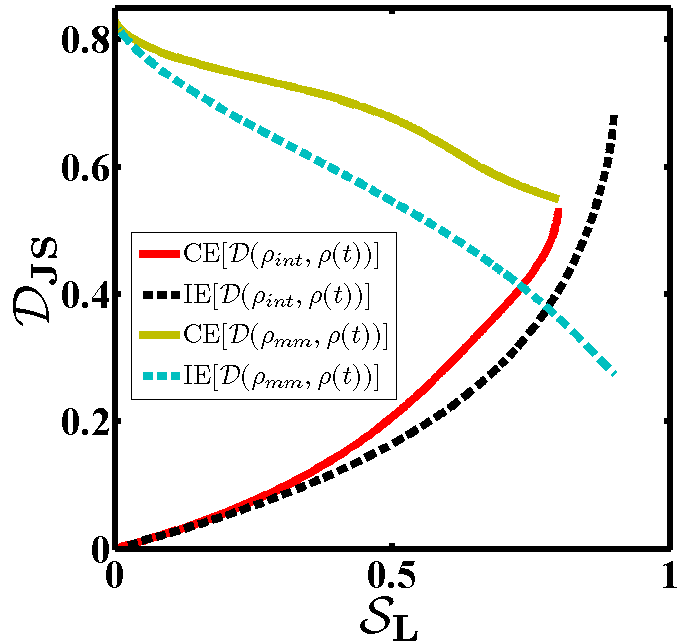}}&  
\subfigure[]{\includegraphics[width=0.22\textwidth]{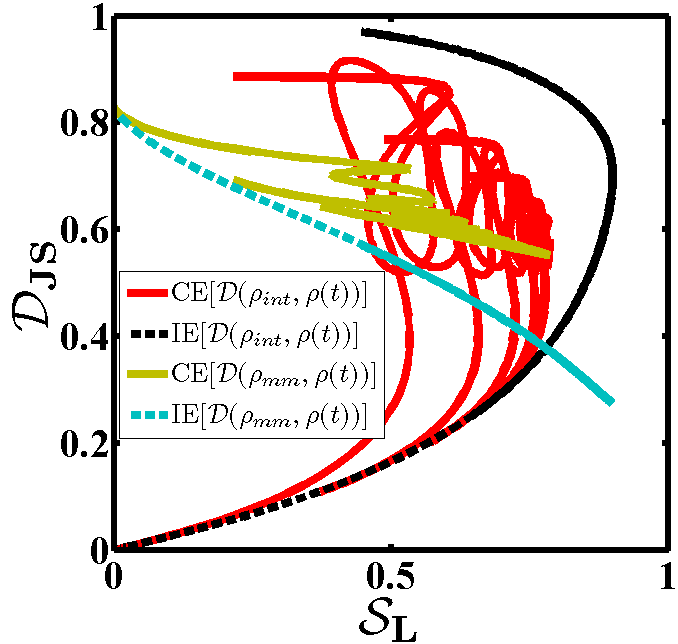}}\\ 
\hline 
\end{tabular}}
\caption{Upper panels: QJSD between the decohered state and: (i) the initial state $ \rho_{int} $ for the cases of IEs (dotted black line) and CE (the solid red line) coupling, (ii) the maximally mixed state $ \rho_{mm} $ also for the cases of IEs (dotted green line) and CE (the solid brown line) coupling, as a function of the linear entropy, in the Markovian (a) and non-Markovian (b) regime. Lower panels: same as in the upper panels for an initial pure W state.}
\label{f11}
\end{figure}

As we have already mentioned in the introduction, in order to investigated the state-space trajectories associated to the final decohered GHZ and W state, we compute the distances between both states and some reference states namely the initial and the maximally mixed states, by recourse to the QJSD. The resultant distances between the mixed state obtained at each time step of the evolution of the initial state of the qubits and the reference states (initial and maximally mixed states) as a function of the linear entropy in the Markovian and non-Markovian regime are reported in Fig.~\ref{f11}, for the case of CE and IEs coupling. 
In Fig.~\ref{f11}, one can readily observe in the Markovian regime that the QJSD between the state of the system affected by RTN and the initial state increases with the increasing of the linear entropy and then freeze to a stationary value when the linear entropy hits its maximum. The opposite is found for the QJSD between the state of the system affected by RTN and the maximally mixed state, where the increases of the degree of mixedness results in the decreasing of the QJSD. These behaviours could be explained by noticing that the initial state of a quantum system gradually evolves towards a mixed state as the alluded system undergoes decoherence. On the other hand, it can also clearly be seen from this figure that the final time-evolved state of the system obtained at the end of the process of decoherence is more mixed in the IEs coupling than in the CE coupling. However, it is worth nothing that the fact that the distance between final time-evolved state of the system and the maximally mixed state is not cancelled clearly shows that the final state the system does not reach a maximally mixed state. Moreover, in the non-Markovian regime, we observe that when the qubits are embedded in a common environment, the trajectories of the system here quantified by the QJSD between the final time-evolved state $ \rho(t) $ and either $ \rho_{int} $ or  $ \rho_{mm} $ are chaotic both for GHZ and W state. Such strange trajectories are connected to the non-Markovian character of the RTN. This figure also reveals once more in the non-Markovian regime that when the qubits are coupled to the noise in independent environments, the trajectory followed by the system has two branches: a dotted line branch and a solid line branch. The former is connected to the initial phase of the entanglement decay and the latter to the entanglement revivals. Indeed, during the entanglement revivals, the system travels partially back and forth on the same curve \cite{35,36}.

\section{Conclusions}\label{s5}
In this paper, we have investigate the dynamics of entanglement and explore some geometrical features of the trajectories in state space, in a system of four non-interacting qubits initially prepared in the GHZ- or W-type states, and coupled to a classical random telegraph noise (RTN) in common and independent environments. The time-evolved state of the system has been obtained by performing the ensemble average over all the possible realizations of the stochastic noise phase picked up by the system during its evolution. We quantify the entanglement by recourse to the negativity, lower bound to concurrence (LBC) and the concept of entanglement witnesses for four-qubit systems. On the other hand, the trajectories of the system in the state space have been characterized by computing by means of the quantum Jensen Shannon divergence (QJSD) the distance between the time-evolved state of the qubits and some reference states, namely the initial and the maximally mixed states. 
\par
Our results show that the evolution of entanglement is drastically affected not only by the nature of system-environment coupling but also by the input configuration of the qubits as well as the Markovian or non-Markovian character of the RTN. Indeed, in the Markovian regime, we found that entanglement is a monotonous function of time while in the non-Markovian regime it is an oscillating function of time. Beyond this, we found that the entanglement of the GHZ-type states is more robust under this noise than the one of the W-type states and that when the qubits are coupled to the noise in a CE, the entanglement survives to the detrimental effects of this last even at sufficiently long time. In particular, for GHZ-type states, we found that the RTN is more fatal to the survival of entanglement existing between the qubits when they are coupled in IEs than when they are embedded in a CE. Stated another way, we found that for an initial GHZ-type state, the CE coupling preserves better the entanglement with respect to the IEs one. However, we found that this is not always true for W-type states. In fact, depending on the purity of the initial state, we found that the IEs coupling may preserve either only for a period of time or indefinitely better the entanglement initially shared between the qubits than the CE coupling. Having compared our results with those of previous investigations, we found that whichever the input state of the qubits and the Q-E configuration (coupling) considered, the three-qubit form of our model \cite{27,28} is less robust (in terms of entanglement preservation) than its four-qubit form (studied in this paper), that is, in the model investigated, the entanglement robustness increases with the increase of the number of qubits of the system. This is a good evidence of the relevance of our model for the development of quantum technologies based on the manipulation of entanglement. Furthermore, even if the ability of the witnesses to detects the presence of entanglement is always weaker than that of the negativity and LBC, we found that for an initial GHZ-type state with $ q>0.821 $, the long-lived entanglement can be successfully detect by the witness operator when the qubits are embedded in a CE. Moreover, depending on the memory properties of the RTN, we found that the trajectories followed by the system may be curvilinear or chaotic and that some features of entanglement dynamics, namely the ESD, ER and the long-lived entanglement can also be identified if we analyse the trajectory of the system in the mixedness-negativity plane. Finally, the influence of the switching rate on the time evolution of entanglement has also been examined and we found that two opposite scenarios may occur depending on whether the qubits are weakly or strongly coupled to the external noise.  Indeed, we found that when the qubits are weakly coupled to the noise, the speed of entanglement decay decreases with the increasing of the bistable fluctuator's switching rate.
 
\begin{widetext}
\begin{appendices}
\renewcommand{\theequation}{\thesection.\arabic{equation}}
\renewcommand{\thesubsection}{\thesection.\arabic{subsection}}
\setcounter{equation}{0}
\section{Explicit forms of the time-evolved four-qubit density matrices: the case of GHZ-type states}\label{A}
Here, we present the explicit forms of the time-evolved density matrices of the system obtained from Eq.~\eqref{a}, when the qubits are prepared in the GHZ-type states as in Eq.~\eqref{5}, for the case of common and independent environments.
\subsection{Independent environments}
For this configuration, we find that the density matrix of the system at a given time $ t $ takes the following form:

\begin{equation}\label{A1}
\begin{split}
\rho_{GHZ_{4}}^{IE}(t)&=\varphi(t)\Big( \lvert 0000\rangle\langle 0000\rvert+\lvert 1111\rangle\langle 1111\rvert \Big)+\phi(t)\Big( \lvert 0000\rangle\langle 1111\rvert+\lvert 1111\rangle\langle 0000\rvert \Big)+\in(t)\Big( \lvert 0001\rangle\langle 0001\rvert+ \lvert 0010\rangle\langle 0010\rvert + 
\\&+
\lvert 0100\rangle\langle 0100\rvert + 
\lvert 0111\rangle\langle 0111\rvert +\lvert 1000\rangle\langle 1000\rvert +  \lvert 1011\rangle\langle 1011\rvert+ \lvert 1101\rangle\langle 1101\rvert +\lvert 1110\rangle\langle 1110\rvert \Big)+
\chi(t)\Big( \lvert 0011\rangle\langle 0011\rvert+ 
\\&+
\lvert 0101\rangle\langle 0101\rvert + \lvert 0110\rangle\langle 0110\rvert + \lvert 1001\rangle\langle 1001\rvert +\lvert 1010\rangle\langle 1010\rvert +  \lvert 1100\rangle\langle 1100\rvert\Big)+
\psi(t)\Big(\lvert 0001\rangle\langle 1110\rvert+\lvert 0010\rangle\langle 1101\rvert +
\\&+
\lvert 0100\rangle\langle 1011\rvert+\lvert 0111\rangle\langle 1000\rvert+\lvert 1000\rangle\langle 0111\rvert+\lvert 1011\rangle\langle 0100\rvert+\lvert 1101\rangle\langle 0010\rvert+\lvert 1110\rangle\langle 0001\rvert\Big)+
\xi(t)\Big(\lvert 0011\rangle\langle 1100\rvert+
\\&+
\lvert 0101\rangle\langle 1010\rvert+
\lvert 0110\rangle\langle 1001\rvert+
\lvert 1001\rangle\langle 0110\rvert+\lvert 1010\rangle\langle 0101\rvert+\lvert 1100\rangle\langle 0011\rvert\Big),
\end{split}
\end{equation}
where 
\begin{equation*}
\begin{split}
&\varphi(t)=\dfrac{1}{16}+\dfrac{3q}{8}\beta_{2}^{2}(t)+\dfrac{q}{16}\beta_{2}^{4}(t);\,\,\in(t)=-\dfrac{q}{16}\beta_{2}^{4}(t)+\dfrac{1}{16};\,\,\chi(t)=\dfrac{1}{16}+q\Big[ \dfrac{1}{16}\beta_{2}^{4}(t)-\dfrac{1}{8}\beta_{2}^{2}(t)\Big];\,\, \phi(t)=\dfrac{q}{16}\Big[ \beta_{2}^{4}(t)+6\beta_{2}^{2}(t)+1\Big];\\& \psi(t)=-\dfrac{q}{16}\Big[ \beta_{2}^{4}(t)-1\Big]\,\, \text{and} \,\, \xi(t)=\dfrac{q}{16}\Big( 1+\beta_{2}(t)\Big)^{2} \Big( \beta_{2}^{2}(t)-2\beta_{2}(t)+1\Big).
\end{split}
\end{equation*}
\subsection{Common environment}
Finally, for the case of CE coupling, the density matrix describing the evolution of the system results into:
\begin{equation}\label{A2}
\rho_{GHZ_{4}}^{CE}(t)=\left[ \begin{array}{cccccccccccccccc}
\varphi(t) & 0 & 0 & \mu(t) & 0 & \mu(t) & \mu(t) & 0 & 0 & \mu(t) & \mu(t) & 0 & \mu(t) & 0 & 0 &  \phi(t)\\ 
0 & \in(t) & \delta(t) & 0 & \delta(t) & 0 & 0 & \delta(t) & \delta(t) & 0 & 0 & \delta(t) & 0 & \delta(t) & \delta(t) &  0\\ 
0& \delta(t) & \in(t) & 0 & \delta(t) & 0 & 0 & \delta(t) & \delta(t) & 0 & 0 & \delta(t) & 0 & \delta(t) & \delta(t) & 0 \\ 
\mu(t) & 0 & 0 & \theta(t) & 0 & \omega(t) & \omega(t) & 0 & 0 & \omega(t) & \omega(t) & 0 & \omega(t) & 0 & 0 & \mu(t) \\ 
0 & \delta(t) & \delta(t) & 0 & \in(t) & 0 & 0 & \delta(t) & \delta(t) & 0 & 0 & \delta(t) & 0 & \delta(t) & \delta(t) & 0 \\ 
\mu(t) & 0 & 0 & \omega(t) & 0 & \theta(t) & \omega(t) & 0 & 0 & \omega(t) & \omega(t) & 0 & \omega(t) & 0 & 0 &  \mu(t)\\ 
\mu(t) & 0 & 0 & \omega(t) & 0 & \omega(t) & \theta(t) & 0 & 0 & \omega(t) & \omega(t) & 0 & \omega(t) & 0 & 0 & \mu(t) \\ 
0 & \delta(t) & \delta(t) & 0 & \delta(t) & 0 & 0 & \in(t) & \delta(t) & 0 & 0 & \delta(t) & 0 &\delta(t)  & \delta(t) & 0 \\ 
0 & \delta(t) & \delta(t) & 0 & \delta(t) & 0 & 0 & \delta(t) & \in(t) & 0 & 0 & \delta(t) & 0 & \delta(t) & \delta(t) & 0 \\ 
\mu(t) & 0 & 0 & \omega(t) & 0 & \omega(t) & \omega(t) & 0 & 0 & \theta(t) & \omega(t) & 0 & \omega(t) & 0 & 0 & \mu(t) \\ 
\mu(t) & 0 & 0 & \omega(t) & 0 & \omega(t) & \omega(t) & 0 & 0 & \omega(t) & \theta(t) & 0 & \omega(t) & 0 & 0 & \mu(t) \\ 
0 & \delta(t) & \delta(t) & 0 & \delta(t) & 0 & 0 & \delta(t) & \delta(t) & 0 & 0 & \in(t) & 0 & \delta(t) & \delta(t) & 0 \\ 
\mu(t) & 0 & 0 & \omega(t) & 0 & \omega(t) & \omega(t) & 0 & 0 & \omega(t) & \omega(t) & 0 & \theta(t) & 0 & 0 & \mu(t) \\ 
0 & \delta(t) & \delta(t) & 0 & \delta(t) & 0 & 0 & \delta(t) & \delta(t) & 0 & 0 & \delta(t) & 0 & \in(t) & \delta(t) & 0 \\ 
0 & \delta(t) & \delta(t) & 0 & \delta(t) & 0 & 0 & \delta(t) & \delta(t) & 0 & 0 & \delta(t) & 0 & \delta(t) & \in(t) & 0 \\ 
\phi(t) & 0 & 0 & \mu(t) & 0 & \mu(t) & \mu(t) & 0 & 0 & \mu(t) & \mu(t) & 0 & \mu(t) & 0 & 0 & \varphi(t)
\end{array} \right]
\end{equation}
where
\begin{equation*}
\begin{split}
&\varphi(t)=\dfrac{15}{64}q+\dfrac{1}{16}+q\left[\dfrac{3}{16}\beta_{4}(t)+\dfrac{1}{64}\beta_{8}(t)\right];\,\,\mu(t)=q\left[\dfrac{1}{16}\beta_{4}(t)+\dfrac{1}{64}\beta_{8}(t)-\dfrac{5}{64}\right];\,\,\phi(t)=q\left[\dfrac{3}{16}\beta_{4}(t)+\dfrac{1}{64}\beta_{8}(t)+\dfrac{19}{64}\right];\\&
\in(t)=\dfrac{1}{16}-q\left[ \dfrac{3}{64}+\dfrac{1}{64}\beta_{8}(t)\right];\,\,\delta(t)=\dfrac{q}{64}\left[ 1-\beta_{8}(t)\right];\,\,\theta(t)=\dfrac{1}{16}-\dfrac{q}{64}+q\left[ -\dfrac{1}{16}\beta_{4}(t)+\dfrac{1}{64}\beta_{8}(t)\right]\,\, \text{and}\\&\omega(t)=q\left[ \dfrac{1}{64}\beta_{8}(t)-\dfrac{1}{16}\beta_{4}(t)+\dfrac{3}{64}\right].
\end{split}
\end{equation*}
The density matrix elements for both $ \rho_{GHZ_{4}}^{IE}(t) $ and $ \rho_{GHZ_{4}}^{CE}(t) $ are written as function of $ \beta_{\kappa}(t) $ $ (\kappa\in \mathbb{N}) $ defined as:
\begin{equation}\label{A3}
\beta_{\kappa}(t)=\Big\langle\cos\left( \kappa\eta_{Q}(t)\right) \Big\rangle_{\eta_{Q}}=
\left\lbrace
\begin{split}
&\displaystyle e^{\displaystyle-\gamma t}\left[ \cosh(\Gamma t)+\dfrac{\gamma}{\Gamma}\sinh(\Gamma t)\right] \rightarrow\,\gamma>\kappa\nu,\,\,\Gamma=\sqrt{\gamma^{2}-(\kappa\nu)^{2}}
\\\\&
\displaystyle e^{\displaystyle-\gamma t}\left[ \cos(\Gamma t)+\dfrac{\gamma}{\Gamma}\sin(\Gamma t)\right] \rightarrow\,\gamma<\kappa\nu,\,\,\Gamma=\sqrt{(\kappa\nu)^{2}-\gamma^{2}}
\end{split},
\right.
\end{equation}
We observe immediately that the form of the time-evolved density matrix of the qubits depends on whether they are coupled to the noise in the common or independent environment.  
\end{appendices}

\begin{appendices}
\renewcommand{\theequation}{\thesection.\arabic{equation}}
\renewcommand{\thesubsection}{\thesection.\arabic{subsection}}
\setcounter{equation}{0}
\section{Explicit forms of the time-evolved four-qubit density matrices: the case of W-type states}\label{B}
Here, we present the explicit forms of the time-evolved density matrices of the system obtained from Eq.~\eqref{a}, when the qubits are prepared in the W-type states as defined in Eq.~\eqref{6}, for the case of common and independent environments. 
\subsection{Independent environments}
For this input configuration, we find that when the qubits are coupled to the RTN in independent environments, the time-evolved density matrix of the system takes the form:
\begin{equation}\label{B1}
\rho_{W_{4}}^{IE}(t)=\left[ \begin{array}{cccccccccccccccc}
\varphi(t) & 0 & 0 & \mu(t) & 0 & \mu(t) & \mu(t) & 0 & 0 & \mu(t) & \mu(t) & 0 & \mu(t) & 0 & 0 & 0\\ 
0 & \in(t) & \delta(t) & 0 & \delta(t) & 0 & 0 & \phi(t) & \delta(t) & 0 & 0 & \phi(t) & 0 & \phi(t) & 0 &  0\\ 
0& \delta(t) & \in(t) & 0 & \delta(t) & 0 & 0 & \phi(t) & \delta(t) & 0 & 0 & \phi(t) & 0 & 0 & \phi(t) & 0 \\ 
\mu(t) & 0 & 0 & \theta(t) & 0 & \omega(t) & \omega(t) & 0 & 0 & \omega(t) & \omega(t) & 0 & 0 & 0 & 0 & \psi(t)\\ 
0 & \delta(t) & \delta(t) & 0 & \in(t) & 0 & 0 & \phi(t) & \delta(t) & 0 & 0 & 0 & 0 & \phi(t) & \phi(t) & 0 \\ 
\mu(t) & 0 & 0 & \omega(t) & 0 & \theta(t) & \omega(t) & 0 & 0 & \omega(t) & 0 & 0 & \omega(t) & 0 & 0 &  \psi(t)\\ 
\mu(t) & 0 & 0 & \omega(t) & 0 & \omega(t) & \theta(t) & 0 & 0 & 0 & \omega(t) & 0 & \omega(t) & 0 & 0 & \psi(t) \\ 
0 & \phi(t) & \phi(t) & 0 & \phi(t) & 0 & 0 & \eta(t) & 0 & 0 & 0 & \chi(t) & 0 &\chi(t)  & \chi(t) & 0 \\ 
0 & \delta(t) & \delta(t) & 0 & \delta(t) & 0 & 0 & 0 & \in(t) & 0 & 0 & \phi(t) & 0 & \phi(t) & \phi(t) & 0 \\ 
\mu(t) & 0 & 0 & \omega(t) & 0 & \omega(t) & 0 & 0 & 0 & \theta(t) & \omega(t) & 0 & \omega(t) & 0 & 0 & \psi(t) \\ 
\mu(t) & 0 & 0 & \omega(t) & 0 & 0 & \omega(t) & 0 & 0 & \omega(t) & \theta(t) & 0 & \omega(t) & 0 & 0 & \psi(t) \\ 
0 & \phi(t) & \phi(t) & 0 & 0 & 0 & 0 & \chi(t) & \phi(t) & 0 & 0 & \eta(t) & 0 & \phi(t) & \phi(t) & 0 \\ 
\mu(t) & 0 & 0 & 0 & 0 & \omega(t) & \omega(t) & 0 & 0 & \omega(t) & \omega(t) & 0 & \theta(t) & 0 & 0 & \psi(t) \\ 
0 & \phi(t) &0 & 0 & \phi(t) & 0 & 0 & \chi(t) & \phi(t) & 0 & 0 & \chi(t) & 0 & \eta(t) & \chi(t) & 0 \\ 
0 & 0 & \phi(t) & 0 & \phi(t) & 0 & 0 & \chi(t) & \phi(t) & 0 & 0 & \chi(t) & 0 & \chi(t) & \eta(t) & 0 \\ 
0 & 0 & 0 & \psi(t) & 0 & \psi(t) & \psi(t) & 0 & 0 & \psi(t) & \psi(t) & 0 & \psi(t) & 0 & 0 & \alpha(t)
\end{array} \right],
\end{equation}
where
\begin{equation*}
\begin{split}
&\varphi(t)=\dfrac{1}{16}+q\left[ \dfrac{1}{8}\beta_{2}(t)-\dfrac{1}{8}\beta_{2}^{3}(t)-\dfrac{1}{16}\beta_{2}^{4}(t)\right];\,\mu(t)=-\dfrac{q}{32}\left[ \beta_{2}^{4}(t)+2\beta_{2}^{3}(t)-2\beta_{2}(t)-1\right];\,\in(t)=\dfrac{1}{16}+\dfrac{q}{16}\left[\beta_{2}(t)+\beta_{2}^{3}(t)+\beta_{2}^{4}(t)\right];
\\& 
\eta(t)=\dfrac{1}{16}+\dfrac{q}{16}\left[\beta_{2}^{4}(t)-\beta_{2}^{3}(t)-\beta_{2}(t) \right];\,\delta(t)=\dfrac{q}{32}\left[ 1+\beta_{2}^{4}(t)+2\beta_{2}^{3}(t)+2\beta_{2}^{2}(t)+2\beta_{2}(t)\right];\,\phi(t)=\dfrac{q}{32}\left[ 1+\beta_{2}^{4}(t)-2\beta_{2}^{2}(t)\right];
\\&
\chi(t)=\dfrac{q}{32}\left[ 1+\beta_{2}^{4}(t)-2\beta_{2}^{3}(t)+2\beta_{2}^{2}(t)-2\beta_{2}(t)\right];
\,\omega(t)=\dfrac{q}{32}\left[1-\beta_{2}^{4}(t)\right];\,
\psi(t)=-\dfrac{q}{32}\left[ \beta_{2}^{4}(t)-2\beta_{2}^{3}(t)+2\beta_{2}(t)-1\right];
\\& 
\theta(t)=\dfrac{1}{16}\left[ 1-q\beta_{2}^{4}(t)\right]\,\text{and}\,\alpha(t)=\dfrac{1}{16}-\dfrac{q}{16}\left[ \beta_{2}^{4}(t)-2\beta_{2}^{3}(t)+2\beta_{2}(t)\right]. 
\end{split}
\end{equation*}

\subsection{Common environment}
On the other hand, when the subsystems are coupled to the RTN in a common environment, the time-evolved density matrix of the system takes the form:
\begin{equation}\label{B2}
\rho_{W_{4}}^{CE}(t)=\left[ \begin{array}{cccccccccccccccc}
\varphi(t) & 0 & 0 & \mu(t) & 0 & \mu(t) & \mu(t) & 0 & 0 & \mu(t) & \mu(t) & 0 & \mu(t) & 0 & 0 & \zeta(t)\\ 
0 & \in(t) & \delta(t) & 0 & \delta(t) & 0 & 0 & \phi(t) & \delta(t) & 0 & 0 & \phi(t) & 0 & \phi(t) & \phi(t) &  0\\ 
0& \delta(t) & \in(t) & 0 & \delta(t) & 0 & 0 & \phi(t) & \delta(t) & 0 & 0 & \phi(t) & 0 & \phi(t) & \phi(t) & 0 \\ 
\mu(t) & 0 & 0 & \theta(t) & 0 & \omega(t) & \omega(t) & 0 & 0 & \omega(t) & \omega(t) & 0 & \omega(t) & 0 & 0 & \psi(t)\\ 
0 & \delta(t) & \delta(t) & 0 & \in(t) & 0 & 0 & \phi(t) & \delta(t) & 0 & 0 & \phi(t) & 0 & \phi(t) & \phi(t) & 0 \\ 
\mu(t) & 0 & 0 & \omega(t) & 0 & \theta(t) & \omega(t) & 0 & 0 & \omega(t) & \omega(t) & 0 & \omega(t) & 0 & 0 &  \psi(t)\\ 
\mu(t) & 0 & 0 & \omega(t) & 0 & \omega(t) & \theta(t) & 0 & 0 & \omega(t) & \omega(t) & 0 & \omega(t) & 0 & 0 & \psi(t) \\ 
0 & \phi(t) & \phi(t) & 0 & \phi(t) & 0 & 0 & \eta(t) & \phi(t) & 0 & 0 & \chi(t) & 0 &\chi(t)  & \chi(t) & 0 \\ 
0 & \delta(t) & \delta(t) & 0 & \delta(t) & 0 & 0 & \phi(t) & \in(t) & 0 & 0 & \phi(t) & 0 & \phi(t) & \phi(t) & 0 \\ 
\mu(t) & 0 & 0 & \omega(t) & 0 & \omega(t) & \omega(t) & 0 & 0 & \theta(t) & \omega(t) & 0 & \omega(t) & 0 & 0 & \psi(t) \\ 
\mu(t) & 0 & 0 & \omega(t) & 0 & \omega(t) & \omega(t) & 0 & 0 & \omega(t) & \theta(t) & 0 & \omega(t) & 0 & 0 & \psi(t) \\
0 & \phi(t) & \phi(t) & 0 & \phi(t) & 0 & 0 & \chi(t) & \phi(t) & 0 & 0 & \eta(t) & 0 & \phi(t) & \phi(t) & 0 \\ 
\mu(t) & 0 & 0 & \omega(t) & 0 & \omega(t) & \omega(t) & 0 & 0 & \omega(t) & \omega(t) & 0 & \theta(t) & 0 & 0 & \psi(t) \\ 

0 & \phi(t) &\phi(t)& 0 & \phi(t) & 0 & 0 & \chi(t) & \phi(t) & 0 & 0 & \chi(t) & 0 & \eta(t) & \chi(t) & 0 \\ 
0 & \phi(t) & \phi(t) & 0 & \phi(t) & 0 & 0 & \chi(t) & \phi(t) & 0 & 0 & \chi(t) & 0 & \chi(t) & \eta(t) & 0 \\ 
\zeta(t) & 0 & 0 & \psi(t) & 0 & \psi(t) & \psi(t) & 0 & 0 & \psi(t) & \psi(t) & 0 & \psi(t) & 0 & 0 & \alpha(t)
\end{array} \right],
\end{equation}
with
\begin{equation*}
\begin{split}
&\varphi(t)=\dfrac{3}{32}q+\dfrac{1}{16}+\dfrac{q}{8}\left[\beta_{2}(t)-\beta_{4}(t)-\beta_{6}(t)-\dfrac{1}{4}\beta_{8}(t)\right];\,\mu(t)=\dfrac{q}{32}\left[-2\beta_{6}(t)-\beta_{8}(t)+2\beta_{2}(t)+1\right];\,\theta(t)=\dfrac{1}{16}-\dfrac{q}{32}\left[1+\beta_{8}(t) \right];
\\&
\in(t)=\dfrac{1}{16}+\dfrac{q}{32}\left[2\beta_{2}(t)+2\beta_{6}(t)+\beta_{8}(t)+\beta_{4}(t)\right];\,\delta(t)=\dfrac{q}{32}\left[2\beta_{2}(t)+2\beta_{6}(t)+\beta_{4}(t)+\beta_{8}(t)+2\right];\,\phi(t)=\dfrac{q}{32}\left[\beta_{8}(t)-\beta_{4}(t)\right];
\\&
\omega(t)=-\dfrac{q}{32}\left[ \beta_{8}(t)-1\right];\,\psi(t)=\dfrac{q}{32}\left[-2\beta_{2}(t)-\beta_{8}(t)+2\beta_{6}(t)+1\right];\,\eta(t)=\dfrac{1}{16}+\dfrac{q}{32}\left[\beta_{8}(t)+\beta_{4}(t)-2\beta_{6}(t)-2\beta_{2}(t)\right];
\\&
\chi(t)=-\dfrac{q}{32}\left[2\beta_{2}(t)-\beta_{4}(t)-\beta_{8}(t)+2\beta_{6}(t)-2\right];\,\alpha(t)=\dfrac{3q}{32}+\dfrac{1}{16}-\dfrac{q}{32}\left[\beta_{8}(t)-4\beta_{6}(t)+4\beta_{4}(t)+4\beta_{2}(t)\right]\,\text{and}
\\&
\zeta(t)=-\dfrac{q}{32}\left[3-4\beta_{4}(t)+\beta_{8}(t)\right].
\end{split}
\end{equation*}
We remark that both density matrices have different forms and can also be written in terms of the time dependent function $ \beta_{\kappa}(t) $ defined in Eq.~\eqref{A3}.

\end{appendices}
\end{widetext}

\end{document}